\documentclass{appolb}
\usepackage{graphicx}
\usepackage{epsfig}
\usepackage{amssymb}
\newcommand{\beq}{\begin{eqnarray}}
\newcommand{\eeq}{\end{eqnarray}}
\newcommand{\nn}{\nonumber \\}
\newcommand{\es}{& = &}
\newcommand{\rs}{\, = \,}
\newcommand{\ps}{& + &}

\newcommand{\ts}{& \times &}
\newcommand{\nt}{\nn \ts}

\begin{document}
\title{ Non-local structure of renormalized Hamiltonian 
        densities on the light-front hyperplane in space-time }
\author{ Stanis{\l}aw D. G{\l}azek 
\address{ Institute of Theoretical Physics,
University of Warsaw, Poland } }
\maketitle
\begin{abstract}
When canonical Hamiltonians of local quantum field 
theories are transformed using a renormalization
group procedure for effective particles, the
resulting interaction terms are non-local. The
range of their non-locality depends on the
arbitrary parameter of scale, which characterizes
the size of effective particles in terms of the
allowed range of virtual energy changes caused by
interactions. This article describes a generic
example of the non-locality that characterizes
light-front interaction Hamiltonian densities of 
first-order in an effective coupling constant. The 
same non-locality is also related to a relative 
motion wave function for a bound state of two 
particles.
\end{abstract}

\vskip-5in
\hfill {\bf IFT/10/05}
\vskip5.4in

%%%%%%%%%%%%%%%%%%%%%%%%%%%%%%%%%%%%%%%%%%%%%%%%%%%%%%%%%%%%
%%%%%%%%%%%%%%%%%%%%%%% INTRODUCTION %%%%%%%%%%%%%%%%%%%%%%%
%%%%%%%%%%%%%%%%%%%%%%%%%%%%%%%%%%%%%%%%%%%%%%%%%%%%%%%%%%%%

\section{ Introduction }
\label{sec:I}

Canonical quantum field theory (QFT) can be
formulated using different forms of Hamiltonian
dynamics~\cite{Dirac}. One of the forms is
distinguished from others by having 7 kinematical
symmetries instead of usual 6. The distinguished
form is called a front form because the
Hamiltonian density in it is defined on a
space-time hyper-plane that is swept by the wave
front of a plane wave of light. The hyper-plane is
called light front (LF). The 7th symmetry is
invariance with respect to the Lorentz boosts
along the direction of motion of the wave. The
additional symmetry has many consequences. For
example, one can describe the relative motion of
constituents of a bound state for arbitrary motion
of the bound state as a whole, achieving a
connection between the rest frame image of the
system with its image in the infinite momentum
frame, and the ground state (vacuum) problem in a
theory is posed in a new way because there is no
spontaneous creation of particles from empty
space~\cite{KogutSusskind}. A canonical
formulation of the standard model (SM) using LF
hyperplane for quantization of fields can be found
in~\cite{Srivastava}.

Canonical Hamiltonians of local theories are
singular operators. They require regularization
and renormalization. In particular, a
renormalization group (RG) procedure produces
Hamiltonians that are called renormalized.
Renormalized Hamiltonians depend on a RG scale
parameter, here called $\lambda$, and are no
longer local. The size of their non-locality
corresponds to the size of $\lambda$. A
renormalized Hamiltonian of some scale $\lambda$
can also be called effective up to the scale
$\lambda$, since it is equivalent to the
Hamiltonian of the original theory and provides an
optimal setup for calculating observables that
concern phenomena whose description does not
require a resolution of effects beyond the
resolution implied by the size of $\lambda$.
Effective Hamiltonians do not depend on
regularization but they do depend on the scale
$\lambda$. This article discusses lowest-order
non-localities that characterize effective
Hamiltonian densities on the LF. Besides general
interest in effective non-local interaction
densities in quantum field theory, LF
non-localities are intriguing because of their
necessarily relativistic nature.

Consider the case of QCD, which is a part of the
SM. The LF power-counting for Hamiltonian
densities in QCD indicates~\cite{Wilsonetal} that
a large number of operators may contribute to a
renormalized Hamiltonian. The power-counting leads
to a complex set of operators because one works in
the Minkowski space, the direction of $z$-axis
(the direction of motion of the plane wave of
light that defines the LF is conventionally chosen
to be against $z$-axis) and the directions
transverse to $z$-axis are treated differently,
and, ab initio, one has to deal with operators
that have important matrix elements arbitrarily
far off energy shell. In summary, effective LF
Hamiltonians are not expected to have a structure
that can be guessed easily. One needs a method to
derive them.

The method discussed here is the renormalization
group procedure for effective
particles~\cite{RGPEP}. The name of the method is
abbreviated RGPEP. The method assumes that finite
parts of counterterms can be fixed using
predictions for observables that follow from the
calculated effective Hamiltonians. The observables
may include properties of bound states even if a
procedure for calculating the Hamiltonians is
itself carried out only in an order-by-order
expansion in powers of an effective coupling
constant. This can be seen on the example of a
Coulomb interaction in the Schr\"odinger
Hamiltonian for an electron and a proton. The
interaction is proportional to the first power of
$\alpha$ and yet it predicts properties of the
electron-proton bound states whose wave functions
have no expansion in powers of $\alpha$ around 0.
Thus, one may hope to learn about non-perturbative
solutions of a theory by solving eigenvalue
problems for Hamiltonians derived using RGPEP even
in low orders of perturbation theory. 

In the ultraviolet regime of QCD, one obtains
non-local vertices that exhibit asymptotic
freedom~\cite{af1,af2} in the dependence of the
effective coupling constant in them, $g_\lambda$,
on the running RG parameter $\lambda$
\cite{gluons}. At the same time, the parameter
$\lambda$ limits from above allowed energy changes
in the interaction vertices\footnote{ In the LF
form of dynamics, $\lambda$ of RGPEP limits
changes in the invariant mass of interacting
particles instead of changes in their energy, and
only a subset of particles that are involved in a
single action of an effective Hamiltonian counts
in the difference.}. This means, through the
connection between energy and momentum and
uncertainty principle for momentum and position
variables, that the vertices resulting from RGPEP
are non-local and the range of the non-locality is
inversely related to the size of $\lambda$. In the
ultraviolet regime, one has $\lambda \gg
\Lambda_{QCD}$, where $\Lambda_{QCD}$ is the
specific momentum scale parameter that physically
characterizes QCD in the scheme of RGPEP. The
limit of infinite $\lambda$, if obtained, could
define a local theory.

At the opposite end of the $\lambda$ scale, there
lies the infrared regime characterized by $\lambda
\sim \Lambda_{QCD}$. In the infrared regime, one
has to deal with Hamiltonian terms that are
involved in the formation of bound states. These
terms are new in the sense that they may
significantly differ in appearance from terms
implied by a classical Lagrangian in a canonical
definition of QCD. In this regime, RGPEP is
designed to guarantee that only invariant mass
changes up to $\lambda \sim \Lambda_{QCD}$ are
allowed to occur (see below). This guarantee is
associated with the characteristic non-localities 
of RGPEP that are described in this article. 

In the infrared regime, one also expects that the
masses of effective particles receive dynamic
contributions order $\Lambda_{QCD}$. Therefore,
the effective dynamics is expected to be limited
to slow relative motion of interacting effective
particles of masses presumably not smaller in size
than about $\Lambda_{QCD}$. 

The infrared regime is not precisely understood.
Nevertheless, on the basis of phenomenology of
hadrons and strong interactions, one expects
formation of constituent quarks and gluons in this
regime and RGPEP is designed to help in
identifying the mechanism that could lead to their
formation as effective particles in QCD. 

RGPEP equations would have to be solved
non-perturbatively in order to provide an exact
effective Hamiltonian of QCD for any $\lambda$.
Exact calculations are hardly possible and it is
not shown that the eigenvalues of LF Hamiltonians,
equivalent to masses squared of physical states,
must be positive. A negative mass squared
eigenvalue would have to be explained. There is no
explanation other than a failure of a theory or a
method applied to seek solutions, such as a wrong
identification of a theory ground state. However,
since the LF RGPEP allows us to calculate
effective Hamiltonians order-by-order in powers of
$g_\lambda$ without making assumptions about the
QCD ground state~\cite{Wilsonetal}, such as the
assumption of vacuum condensates~\cite{SVZ}, the
structure of the LF non-local interactions that
result from RGPEP in the order-by-order
calculations are of interest as elements of the
unknown realm. This article provides a description
of the leading non-localities that show up already
in the first-order terms, independently of
additional non-local effects that build up on top
of the leading ones in higher-order terms.

Interest in the non-locality that characterize
RGPEP is also motivated by a general observation
that the non-locality contributes to a replacement
of an abstract local theory (the one that requires
regularization) by a perhaps more realistic
effective one, whose scale parameter $\lambda$ can
be adjusted in order to obtain a Hamiltonian whose
physically most important terms look simplest
possible \cite{Wilson1965}. In this sense, the
non-locality is obtained on the basis of RG ideas
with no need for invoking any new degrees of
freedom, such as, for example, the ones introduced
in the case of a non-local string picture of
particles\footnote{Questions concerning gravity
and its relation to quantum mechanics require a
separate discussion. Also, conditions of causality
may constrain Hamiltonian non-localities and it is
not known if RGPEP can satisfy them automatically
in gauge theories. The author would like to thank
J. Lukierski for a comment concerning the latter
issue.} or other forms of substructure, perhaps
even including additional symmetries. From this
point of view, the non-local structure of
interactions that RGPEP may lead to is of more
general interest than QCD alone. In fact, the
first-order non-locality discussed in this article
occurs in QFT in the same form irrespective of
many details concerning spin and other quantum
numbers of effective particles, such as isospin,
flavor, or color. For example, it would be the
same in perturbative quantum gravity.

Section \ref{sec:nlvims} reviews qualitative
features of non-local interaction vertices in
lowest-order RGPEP in momentum space. The main
Section \ref{sec:nlvips} discusses non-locality in
space-time in several subsections, starting in
Section \ref{How} from the question of how
non-local interactions become local when $\lambda
\rightarrow \infty$. The non-locality for finite
$\lambda$ and particle mass $m \rightarrow 0$ is
discussed in Section \ref{m->0}. Non-locality for
$\lambda \sim m$ is described in Section
\ref{lambda<m}. A comparison with a
non-relativistic theory is provided in Section
\ref{nr}. Section \ref{wavefunction} explains the
relationship between the relativistic
non-localities and 2-body bound-state wave
functions. Section \ref{c} concludes the paper
with a summary and a few comments. An Appendix in
three parts provides some details of calculations.

%%%%%%%%%%%%%%%%%%%%%%%%%%%%%%%%%%%%%%%%%%%%%%%%%%%%%%%%%%%%
%%%%%%%%%%%%%%%%%%%%%%%  SECTION  1  %%%%%%%%%%%%%%%%%%%%%%%
%%%%%%%%%%%%%%%%%%%%%%%%%%%%%%%%%%%%%%%%%%%%%%%%%%%%%%%%%%%%

\section{ Non-local vertices in momentum space }
\label{sec:nlvims}

Let a regularized Hamiltonian with counterterms
for some QFT be denoted by $H$ and let $a$ denote
creation or annihilation operators of bare
particles in this Hamiltonian. For example, the
bare particles in the case of LF Hamiltonian of
QCD are quanta of canonical quark and gluon fields
in the gauge $A^+= A^0 + A^3$, with some
arbitrarily large cutoffs on momenta. The cutoffs
are meant to be arbitrarily large in the sense of
tending to the limit of removing the
regularization.

In RGPEP, one introduces effective particles of
scale $\lambda$ through a transformation
$U_\lambda$ (see below) whose construction
involves the entire $H$~\cite{RGPEP}. Counterterms
introduced in $H$ allow one to construct a formal
expansion in a series of powers of the effective
coupling constant $g_\lambda$, but it should be
stressed that even the Hamiltonian of order
$g_\lambda^4$ is not fully known yet in important
theories \cite{gluons,GlazekMlynik}. Since the
counterterms are constructed using RGPEP (to
identify their structure) and results for
observables (to fix the unknown finite parts), the
complete transformation $U_\lambda$ can be made
unitary in an order-by-order construction if one
can simultaneously change the initial $H$ by
including counterterms also order-by-order.
However, some characteristic terms can be
introduced partly in a non-perturbative way. 

The key example is provided by a quark
mass-squared term to which one can add a term
proportional to $\Lambda^2_{QCD} \sim \lambda^2
\exp{ (- b/\alpha_\lambda)}$, where $b$ is a
constant and $\alpha_\lambda =
g_\lambda^2/(4\pi)$. Such addition contributes 0
to an expansion in powers of $g_\lambda$ around
$g_\lambda = 0$ but the calculation of $U_\lambda$
can be carried out in RGPEP for arbitrary values
of particle masses. Then, on the one hand, mass
terms can be adjusted by matching theoretical
predictions with observables. On the other hand,
the question of how a non-perturbative dynamics
described by the calculated Hamiltonian relates
mass terms to observables requires investigation.
Other examples of inclusion of non-perturbative
effects in a LF Hamiltonian density are provided
by the arrangement of couplings according to the
rules of coupling coherence that keeps track of
symmetries \cite{PerryWilson}, including the
arrangement of operators allowed by power-counting
so that the resulting set of terms corresponds to
a theory with a spontaneously broken symmetry
\cite{Wilsonetal}.

The example of a mass term in $H_0$ is important
in discussion of non-local Hamiltonian densities
because eigenvalues of $H_0$ are used to define
the RGPEP factors responsible for the
non-locality. The non-locality of effective QCD
depends on the quark and gluon mass parameters.
Since QCD promises to generate contributions to
the effective particle masses from
$\Lambda_{QCD}$, it should be stressed that the
parameter $\Lambda_{QCD}$ primarily characterizes
perturbative $\lambda$-dependence of effective
theories for large $\lambda$. In theories with
large $\lambda$, small mass terms may be treated
as negligible. In this context, the variation of
non-locality of effective LF QCD Hamiltonians over
a large range of $\lambda$ appears related to the
question of generation of particle masses in local
theories with formal symmetries considered valid
for strictly massless particles (chiral symmetry).
This article discusses non-localities for
different ratios of a mass parameter $m$ to the
scale parameter $\lambda$, irrespective of the 
value of parameters like $\Lambda_{QCD}$.

The RGPEP operation $U_\lambda$ mentioned above, 
transforms creation and annihilation operators 
by a rotation,
\beq
\label{blambda}
a_\lambda \es U_\lambda \, a \, U_\lambda^\dagger \, . 
\eeq
The corresponding Hamiltonian operator $H_\lambda$ 
is constructed to be the same as $H$,
\beq
\label{H}
H_\lambda(a_\lambda) \es H(a) \, .
\eeq
Consequently, $H_\lambda(a_\lambda)$ obtainable in
perturbative RGPEP is assumed to be a combination
of products of operators $a_\lambda$ with
coefficients $c_\lambda$ that are different from
coefficients $c$ of corresponding products of
operators $a$ in $H(a)$. RGPEP provides
differential (or algebraic) equations that produce
expressions for the coefficients $c_\lambda$ in
$H_\lambda$~\cite{RGPEP}. Namely, from
\beq 
H_\lambda( a ) 
\es 
U_\lambda^\dagger H(a) \, U_\lambda \, ,
\eeq
and the condition $U_\infty = 1$, one obtains
\beq
{d \over d\lambda } \, H_\lambda( a ) 
\es
[T_\lambda, H_\lambda(a)] \, , \\
H_\infty(a) \es H(a) \, ,
\eeq
where 
\beq
T_\lambda \es 
- U_\lambda^\dagger {d \over d\lambda } \,
U_\lambda \, .
\eeq
Therefore, the evolution of coefficients
$c_\lambda$ with $\lambda$ is determined by
$T_\lambda$; see Eqs. (2.28) and (2.29)
in~\cite{RGPEP}. 

Since $T_\lambda$ vanishes when interactions
vanish, the Hamiltonian $H_\lambda(a)$ can be
expanded in powers of the interaction strength.
Suppose the interaction strength is parameterized
by a suitably chosen coupling constant
$g_\lambda$. In the case of QCD, the limit of
vanishing $g_\lambda$ for any fixed value of
$\lambda$ can also be seen as a limit of vanishing
$\Lambda_{QCD}$. In this limit, the
non-perturbative terms mentioned above,
proportional to positive powers of
$\Lambda_{QCD}$, are also vanishing.

The first-order terms in $H_\lambda(a_\lambda)$,
i.e., terms proportional to the first power of
$g_\lambda$, are functions of operators
$a_\lambda$. As such, they have the same form as
the canonical interaction terms proportional to
the bare coupling constant $g$ have as functions
of the bare operators $a$, containing $c_\infty =
c$. The only difference is that $g$ is replaced by
$g_\lambda$ and $c$ is replaced by $c_\lambda =
f_\lambda c $, where $f_\lambda$ is a vertex form
factor of RGPEP~\cite{RGPEP}. Precisely this form
factor introduces the non-locality studied in this
article.

The first step in studying the non-locality
resulting from the form factor $f_\lambda$ is to
define a generic example of the interaction term
in which $f_\lambda$ appears. For this purpose,
one can observe that in physically important
canonical local theories, such as gauge theories
or Yukawa theory, the Lagrangian interaction
densities of first-order in the coupling constants
contain products of three fields evaluated at the
same space-time point. For example, fermion fields
$\psi$ are coupled with gauge boson fields $A$ in
a product of the type $g \bar \psi /
{\hspace{-7.5pt}} A \psi$, and they are coupled
with scalar fields $\phi$ through the product of
the type $g \bar \psi \phi \psi$. Non-Abelian
gauge fields are coupled to themselves through a
product of the type $g Tr \, \partial_\mu A_\nu
[A^\mu, A^\nu] $. Therefore, for the present
discussion of first-order non-localities, it is
sufficient to consider space-time operator
densities that are products of three fields. 

Another observation is that $f_\lambda$ only
depends on the change of energy across an
interaction Hamiltonian $H_{\lambda I} = H_\lambda
- H_{\lambda 0}$. More precisely, in LF dynamics,
$f_\lambda$ depends on the change of an invariant
mass of the interacting particles across the
interaction.\footnote{At the same time, three
components of a total momentum, $P^+=P^0+P^3$ and
$P^\perp = (P^1, P^2)$, are preserved and the
invariant-mass change is invariant with respect to
7 Poincar\'e transformations that preserve the LF
hyperplane.} This means that the first-order
non-locality structure does not depend on spin,
isospin, flavor or color variables. In other
words, the first-order RGPEP evolution of
coefficients $c_\lambda$ with $\lambda$ is limited
to variation of the range of allowed changes in
the invariant mass and this change depends only on
the momenta and masses of the interacting
particles.

According to these two observations, generic
features of the first-order non-locality can be
studied in the case of a product of three scalar
fields. Conclusions regarding first-order
non-locality in interactions of more complex
fields will be the same as for scalars, except for
additional algebraic factors or derivatives that
originate directly from the vertices of
corresponding canonical theories.

Consider the classical canonical LF interaction 
Hamiltonian for a real (chargeless) scalar field 
$\psi(x)$,
\beq
\label{HI}
H_I \es g \int dx^- d^2 x^\perp \, : \psi^3(x) : \, ,
\eeq
where $x^- = x^0 - x^3$, $x^\perp = (x^1, x^2)$,
and the LF is defined by the condition $x^+=x^0 + 
x^3=0$. Hermitian quantum field $\psi(x)$ is 
composed of creation and annihilation operators 
for bare particles,
\beq
\label{psi}
\psi(x)        \es  \int [p] \, a_p \, e^{ - i p \, x}  
               \rs  \psi^\dagger(x)                     \, , 
\eeq
where
\beq
\label{pintegral}
\int [p]       \es  \int {d^3 p \over 2|p^+| (2\pi)^3 } \, ,  \\
a_{-p}         \es  a^\dagger_p \, ,
\eeq
$d^3p = dp^+ d^2p^\perp$, and all three integrals 
over momentum variables extend from $-\infty$ to 
$+\infty$. Single commutation relation
\beq
{[}a_p, a_q{]} \es  2p^+(2\pi)^3 \delta^3(p+q)          \, ,  
\eeq
contains three corresponding commutation relations: 
one for two creation operators when both $p^+$ and 
$q^+$ are negative, one for two annihilation operators 
when both $p^+$ and $q^+$ are positive, and one for 
one creation operator and one annihilation operator, 
when $p^+$ and $q^+$ have opposite signs. The 
commutation relation implies
\beq
{[}\psi(x) , \partial^+  \psi(y) {]} 
               \es  i \delta^3(x-y)
\, ,
\eeq
where $\partial^+ = 2\partial/\partial x^-$.
One also has
\beq
\label{ap}
a_p            \es  |p^+| \int d^3x \, e^{ + i p
\, x} \, \psi(x) \, , 
\eeq
where $d^3x = dx^- d^2 x^\perp$ and all three integrals 
over position variables extend from $-\infty$ to 
$+\infty$ on the LF.

The above notation differs from the standard
one~\cite{Yan1,Yan2}. The difference is that the
creation and annihilation operators are
distinguished solely by the sign of $p^+$ in $a_p$
and the integration over momentum component $p^+$
is not limited to only positive values. The colon
sign in Eq. (\ref{HI}) denotes normal ordering.
The normal ordering is defined using Feynman's
convention~\cite{Feynman} with the ordering
parameter set equal to $p^+$ that ranges from
$-\infty$ to $+\infty$. In this convention, it is
understood that an operator $a_{p_1}$ stands to
the left of the operator $a_{p_2}$ when $p_1^+ <
p_2^+$. Otherwise, the order is reversed. Thus,
all annihilation operators $a_p$, which by
definition have $p^+>0$, stand to the right of all
creation operators, which by definition are $a_p$
with $p^+<0$. 

In terms of operators $a$, the local, canonical 
interaction Hamiltonian reads
\beq
H_I \es g
\int [p_1 p_2 p_3]                      \,
2 (2\pi)^3 \, \delta^3(p_1 + p_2 + p_3) \,
\,: a_{p_1} \, a_{p_2} \, a_{p_3} :         \, .
\eeq
From this expression, RGPEP produces an effective 
interaction Hamiltonian of first order in the 
form~\cite{RGPEP} 
\beq
\label{HlambdaI}
H_{\lambda I} \es 
g_\lambda 
\int [p_1 p_2 p_3] \,
2 (2\pi)^3 \, \delta^3(p_1 + p_2 + p_3) \,
\, f_\lambda \,\, : a_{\lambda p_1} \, a_{\lambda p_2} \, a_{\lambda p_3} : \, , 
\eeq
where 
\beq
f_\lambda      \es e^{- (\Delta{\cal
M}^2/\lambda^2  )^2} 
\eeq
is the source of non-locality of the effective vertex. 
The argument of the form factor $f_\lambda$ is the
difference between the invariant masses of annihilated 
particles, ${\cal M}_a^2$, and created particles, 
${\cal M}_c^2$. Namely,
\beq
\label{masses}
\Delta{\cal M}^2
\rs
{\cal M}_a^2 - {\cal M}_c^2 \, , 
\quad
{\cal M}_a^2 \rs  P_a^2 \, ,
\quad \quad 
{\cal M}_c^2 \rs  P_c^2 \, ,
\eeq
where the total momentum four-vectors for
annihilated and created particles are
\beq
\label{momenta}
P_a \es \sum_{i=1}^3 \, \theta( p_i^+) \, p_i \, ,
\quad 
P_c \rs \sum_{i=1}^3 \, \theta(-p_i^+) \, p_i \, , 
\quad 
p_i^- \rs {p_i^{\perp 2} + m^2 \over p_i^+ } \, .
\eeq
If a Hamiltonian density term contained $n > 3$ fields, 
the sums over momenta in the above expressions that 
define $f_\lambda$ as function of a change in the 
invariant mass of interacting particles would extend 
up to $n$ instead of only 3, with no other change.

%%%%%%%%%%%%%%%%%%%%%%%%%%%%%%%%%%%%%%%%%%%%%%%%%%%%%%%%%%%%
%%%%%%%%%%%%%%%%%%%%%%%  SECTION  2  %%%%%%%%%%%%%%%%%%%%%%%
%%%%%%%%%%%%%%%%%%%%%%%%%%%%%%%%%%%%%%%%%%%%%%%%%%%%%%%%%%%%

\section{ Non-local vertices in position space }
\label{sec:nlvips}

In the first step of defining effective LF Hamiltonian
densities on the LF, we introduce effective quantum 
field operators. This is done in analogy with Eqs. 
(\ref{psi}) to (\ref{ap}) for bare fields. Namely,
\beq
\label{psil}
\psi_\lambda(x) \es  \int [p] \, a_{\lambda p} \, e^{ - i p \, x}  
                \rs  \psi_\lambda^\dagger(x)                     \, , 
\eeq
where
\beq
\label{al-p}
a_{\lambda \, -p}         \es  a^\dagger_{\lambda
p}  \, , \\
{[}a_{\lambda p}, a_{\lambda q}{]} \es
2p^+(2\pi)^3 \delta^3(p+q) \, ,  \\
\label{alp}
a_{\lambda p}            \es  |p^+| \int d^3x \, e^{ + i p
\, x} \, \psi_\lambda(x) \, . 
\eeq
Thus, the effective field commutation relations remain
the same irrespective of the value of $\lambda$,
\beq
{[}\psi_\lambda(x) , \partial^+  \psi_\lambda(y) {]} 
               \es  i \delta^3(x-y)
\, .
\eeq
Using Eq. (\ref{alp}), the effective interaction 
$H_{\lambda I}$ is obtained in the form
\beq
H_{\lambda I} 
\es 
g_\lambda 
\int [p_1 p_2 p_3] \,
2 (2\pi)^3 \, \delta^3(p_1 + p_2 + p_3) \,
f_\lambda \, |p_1^+ p_2^+ p_3^+|
\\
\ts
\int d^3x_1 \, d^3x_2 \, d^3x_3 \, 
e^{ + i ( p_1 \, x_1 + p_2 \, x_2 + p_3 \, x_3 ) } \, 
: \psi_\lambda(x_1) \, \psi_\lambda(x_2) \, \psi_\lambda(x_3) : \, .
\nonumber 
\eeq
This means that 
\beq
\label{HlambdaIp}
H_{\lambda I} 
\es 
g_\lambda 
\int d^3x_1 \, d^3x_2 \, d^3x_3 \, 
\tilde f_\lambda(x_1, x_2 , x_3) \,
: \psi_\lambda(x_1) \, \psi_\lambda(x_2) \, \psi_\lambda(x_3) : \, ,
\eeq
where the space-time non-locality of the effective
interaction Hamiltonian density on the LF is described 
by the function
\beq
\label{tildef}
\tilde f_\lambda(x_1, x_2 , x_3)
\es
\int [p_1 p_2 p_3] \,
2 (2\pi)^3 \, \delta^3(p_1 + p_2 + p_3) \,
| p_1^+ p_2^+ p_3^+ | \,
\nn
\ts
f_\lambda \, 
e^{ + i (p_1 x_1 + p_2 x_2 + p_3 x_3) } \, .
\eeq
Evaluation of this integral leads to the main 
results concerning non-locality of effective 
LF Hamiltonians. But in order to introduce
relevant concepts we first discuss the issue
of how the non-local interactions become local
when $\lambda \rightarrow 0$.

%%%%%%%%%%%%%%%%%%%%%%%%%%%%%%%%%%%%%%%%%%%%%%%%%%%%%%%%%%%%
%%%%%%%%%%%%%%%%%%%%%%%  SUBSECTION  %%%%%%%%%%%%%%%%%%%%%%%
%%%%%%%%%%%%%%%%%%%%%%%%%%%%%%%%%%%%%%%%%%%%%%%%%%%%%%%%%%%%

\subsection{ How non-local interactions become 
             local when $\lambda \rightarrow \infty$ }
\label{How}

Locality of an interaction Hamiltonian in the
limit of $\lambda \rightarrow \infty$ is here
understood as the following feature: matrix
elements of the interaction Hamiltonian vanish
between states of effective particles
corresponding to RGPEP scale $\lambda$ if the wave
functions of these particles in these states have
supports separated by a distance $r$ that is not
smaller than some fixed but arbitrarily small
distance $r_0$ so that $r\lambda \rightarrow
\infty$. In other words, locality in the limit
$\lambda \rightarrow \infty$ means that matrix
elements between all states that are
experimentally separable in space vanish and all
experimentally accessible separations $r$ are
considered different from 0 by no less than some
very small but fixed amount $r_0$ when $\lambda
\rightarrow \infty$. Every local theory may be
discovered inapplicable in physics when $r_0$ is
reduced below certain value. In such case, $r_0$
designates the limit of physical applicability of
that local theory.

We begin by showing how Eq. (\ref{tildef})
produces an interaction that becomes local when
$\lambda \rightarrow \infty$. Formally, the
notation introduced in Eqs. (\ref{pintegral}),
(\ref{psil}), and (\ref{alp}) produces this result
pointwise in an obvious way: $\lim_{\lambda
\rightarrow \infty} f_\lambda = f_\infty = 1$,
factors $1/|p^+|$ in the integration measures are
cancelled by the factors of $|p^+|$ in the
expressions for $a_p = a_{\infty p} =
\lim_{\lambda \rightarrow \infty} a_{\lambda p}$,
and the remaining integrals produce 
\beq
\tilde f_\infty(x_1, x_2 , x_3)
\es
\delta^3(x_1 - x_3) \, \delta^3(x_2 - x_3) \, .
\eeq
This result leads through Eq. (\ref{HlambdaIp}) to
Eq. (\ref{HI}), with operators $\psi_\infty(x)$
$=$ $\lim_{\lambda \rightarrow \infty}$
$\psi_\lambda(x)$ $=$ $\psi(x)$ as a consequence
of $a$ $=$ $a_\infty$ $=$ $\lim_{\lambda
\rightarrow \infty}$ $a_\lambda$. It is also
understood that $\lim_{\lambda \rightarrow
\infty}$ $g_\lambda$ $=$ $g_\infty$ and $g_\infty$
differs from $g$ only by the value implied by a
coupling constant counterterm. In this reasoning,
the space-time variable $x_3$ in Eq.
(\ref{HlambdaIp}) plays the role of variable $x$
in Eq. (\ref{HI}).

On the other hand, this mechanism appears to
require further explanation because particle
momenta $p^+$ are limited to only positive values
in the standard LF notation for canonical quantum
fields $\psi(x)$. The same standard notation could
be used in the case of effective fields
$\psi_\lambda(x)$, and all integrals over $p^+$
would only extend from 0 to $+\infty$. Instead, in
our notation, the momentum conservation
$\delta$-function in Eq. (\ref{HlambdaI}) forces
one of the momenta $p_1$, $p_2$ and $p_3$ to have
an opposite $+\,$-component to the two
others~\footnote{The case that all three momenta
have components $p^+=0$ is excluded by
regularization in a canonical theory by demanding
that $|p^+|$ in the Fourier expansion of every
field in a Hamiltonian density is greater than
some positive infinitesimal constant $\epsilon^+$.
The strong limit of $\epsilon^+ \rightarrow 0$ is
immediately taken in all terms obtained from the
integration on the LF in the theory regularized
with an infinitesimally small $\epsilon^+$.
Subsequent transformation $U_\lambda$ replaces
operators $a$ with $a_\lambda$ preserving their
momentum labels.}. According to Eq. (\ref{alp}),
this means that the only terms that contribute are
those in which one particle of positive $p^+$ is
annihilated and two particles of positive $p^+$ 
are created, or two are annihilated and one is 
created. There are 3 terms of each type. Therefore, 
the effective interaction equals
\beq 
\label{HlambdaI3} 
H_{\lambda I} 
\es 
3 g_\lambda \, \prod_{i=1}^3 \int_0^\infty {dp^+_i
\over 2p_i^+ (2\pi)}\int {d^2p_i^\perp \over
(2\pi)^2 } \nt 2 (2\pi)^3 \, \delta^3(p_1 + p_2 -
p_3) \, \, f_\lambda \, \left( a_{\lambda
p_1}^\dagger \, a_{\lambda p_2}^\dagger \,
a_{\lambda p_3} + h.c. \right) \, , 
\eeq 
where
\beq 
f_\lambda \es e^{ -({\cal M}_{12}^2 -
m^2)^2/\lambda^4} \, .
\eeq
The sign of normal ordering is not needed. 
Thus, one can write
\beq 
\label{HlambdaI4} 
H_{\lambda I} 
\es 
3 g_\lambda \, \left[ \prod_{i=1}^3 \int_0^\infty {dp^+_i
\over 2 (2\pi)}\int {d^2p_i^\perp \over
(2\pi)^2 } \right] \, 2 (2\pi)^3 \, \delta^3(p_1 + p_2 -
p_3) \, \, f_\lambda \,
\nt
\int d^3x_1 d^3x_2 d^3x_3 \, 
\left[
 \, e^{ - i (p_1 x_1 + p_2 x_2 - p_3 x_3) } \, 
\psi_\lambda(x_1)\psi_\lambda(x_2) \psi_\lambda(x_3)
+ h.c. \right] \, , \nn
\eeq
without normal ordering. It is visible that
the momentum components $p_1^+$ and $p_2^+$ are
all always positive and they sum up to an always
positive $p_3^+$. This implies that one can write
$p_1^+ = z p_3^+$, $p_2^+ = (1-z) p_3^+$, and the
range of integration over $z$ is from 0 to 1.
Moreover, introducing parameterization
\beq
\label{p1standard}
p_1^+     \es z P^+ \, , 
\quad
p_1^\perp \rs z P^\perp + q^\perp \, , 
\quad
p_1^-     \rs { p_1^{\perp 2} + m^2 \over z P^+ } \, , \\
\label{p2standard}
p_2^+     \es (1-z) P^+ \, , 
\quad
p_2^\perp \rs (1-z) P^\perp - q^\perp \, , 
\quad
p_2^-     \rs { p_2^{\perp 2} + m^2 \over (1-z)P^+ } \, , 
\eeq
which is a standard way of parameterizing relative
motion of interacting particles 1 and 2 with
variables $z$ and $q^\perp$ in canonical LF dynamics, 
one obtains the invariant mass squared of particles 
1 and 2 in the argument of the form factor $f_\lambda$ 
equal
\beq
\label{M12}
{\cal M}_{12}^2
\es
{ q^{\perp 2} + m^2 \over z(1-z) } \, .
\eeq
The above relations hold independently of the value 
of $\lambda$. Changing integration variables to
$P^+$, $P^\perp$, $z$, and $q^\perp$, one obtains
for $x^+=0$ that
\beq 
\label{HlambdaI5} 
H_{\lambda I} 
\es 
3 g_\lambda \, 
\int_0^\infty {dP^+       \over 2 (2\pi)   } 
\int          {d^2P^\perp \over   (2\pi)^2 } 
\int_0^1 {dz P^+ \over 2 (2\pi)} \int {d^2 q^\perp \over (2\pi)^2 } 
\, f_\lambda \,
\nt
\int d^3x_1 d^3x_2 d^3x_3 \, 
\left[
 \, e^{ - i X } \, 
\psi_\lambda(x_1)\psi_\lambda(x_2) \psi_\lambda(x_3)
+ h.c. \right] \, , \\
\label{X}
X \es (zP+q) x_1 + [(1-z)P-q] x_2 - P x_3 \, .
\eeq
But this result means that
\beq 
\label{HlambdaI6} 
H_{\lambda I} 
\es 
g_\lambda \,
\int d^3x_1 d^3x_2 d^3x_3 \, 
\bar f_\lambda(x_1, x_2, x_3) \, 
\psi_\lambda(x_1)\psi_\lambda(x_2)
\psi_\lambda(x_3) + h.c. \, ,
\nn
\eeq
where
\beq
\label{barf}
\bar f_\lambda(x_1, x_2, x_3)
\es
3 \int_0^\infty {dP^+       \over 2 (2\pi)   } 
\int          {d^2P^\perp \over   (2\pi)^2 } 
\int_0^1 {dz P^+ \over 2 (2\pi)} \int {d^2 q^\perp \over (2\pi)^2 } 
\, f_\lambda \, e^{ - i X } \, .
\eeq
There is also no need for the sign of normal ordering 
in Eq. (\ref{HlambdaI6}), since the signs of momentum
variables in the integration automatically put
creation operators to the left of annihilation
operators.

Eqs. (\ref{HlambdaI6}) and (\ref{barf}) should be
compared with Eqs. (\ref{HlambdaIp}) and
(\ref{tildef}). The issue to clarify is how the
integration over $z$ only from 0 to 1 and over
only positive values of $P^+$ leads to a local
interaction when $\lambda \rightarrow \infty$. The
limited range of integration in momentum space
seems to always require some smearing of
interaction in position space and the variable $z$
is limited in Eq. (\ref{barf}) to the range from 0
to 1, irrespective of the size of $\lambda$. How
does locality emerge in the limit of $\lambda
\rightarrow \infty$? Details of relevant reasoning
are collected in Appendix \ref{a:connection}. Here
we only provide a description of the connection
between the support of $f_\lambda$ as function of
$z$ between 0 and 1 and the locality of
$H_{\lambda I}$ in the limit $\lambda \rightarrow
\infty$.

The non-locality of interaction Hamiltonian in 
Eq. (\ref{HlambdaIp}) is described by the function 
$\tilde f_\lambda (x_1, x_2, x_3)$ defined in 
Eq. (\ref{tildef}) by an integral over momentum
variables. Eq. (\ref{atildef3}) in Appendix 
\ref{a:connection} shows that
\beq
\label{atildef7main}
\tilde f_\lambda(x_1, x_2 , x_3)
\es
\int { d^3 P                                \over 2 (2\pi)^3} 
\int_{-\infty}^{+\infty} {d\zeta |P^+|      \over 2 (2\pi)  } 
\int {d^2\kappa^\perp                       \over   (2\pi)^2} 
\,
f_\lambda \, e^{ - i Y } \, , 
\\
f_\lambda 
& \equiv & 
f_\lambda(\zeta,\kappa)
\rs
e^{- (\Delta{\cal M}^2/\lambda^2  )^2} \, , \\
\label{DeltaMain}
|\Delta {\cal M}^2 |
\es
\left[ { \kappa^{\perp 2} + m^2 \over
\zeta(1-\zeta) }  - m^2 \right] 
{1 + |\zeta| + |1-\zeta| \over 2}  \, , \\
\label{aY7main}
Y
\es
(\zeta P + \kappa) x_1 + [(1-\zeta) P - \kappa]
x_2 - P x_3 \, .
\eeq
The Hamiltonian obtained by integrating a
normal-ordered product of three quantum fields
$\psi(x_1)$, $\psi(x_2)$, $\psi(x_3)$ with
function $\tilde f_\lambda (x_1, x_2, x_3)$ over
LF hyperplane in Eq. (\ref{HlambdaIp}), is the
same as the Hamiltonian obtained in Eq.
(\ref{HlambdaI6}). In Eq. (\ref{atildef7main}) for
$\tilde f_\lambda (x_1, x_2, x_3)$, integration
over the range of negative $P^+$ corresponds to
the sign of Hermitian conjugation, $h.c.$, in Eq.
(\ref{HlambdaI6}). Integrations over $\zeta < 0$
and $\zeta > 1$ contribute the same operator as
the integration over $\zeta$ between 0 and 1 does.
Thus, the three regions contribute
\begin{figure}
\label{fig:FZetaKappa}
\begin{center}
 \epsfig{file=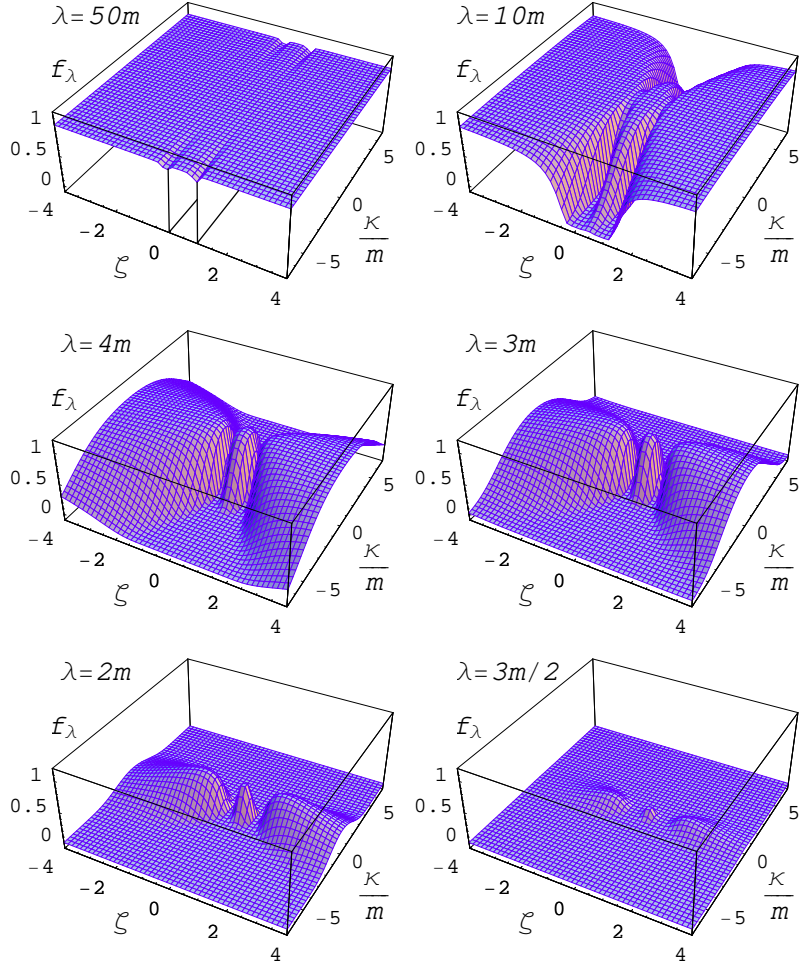,scale=1.35}
\caption{The RGPEP vertex form factor $f_\lambda \equiv 
f_\lambda(\zeta,\kappa)$ that is integrated over momenta 
in Eqs. (\ref{tildef}) and (\ref{atildef7main}) to 
produce the non-local interaction density in 
$H_{\lambda I}$ in Eq. (\ref{HlambdaIp}), for 
six values of $\lambda$. $m$ is the mass of interacting 
particles. The three bumps correspond to the regions 
$\zeta <0$, $0 < \zeta <1$, and $\zeta > 1$, all three 
contributing equally to $H_{\lambda I}$. When $\lambda 
\rightarrow \infty$, $f_\lambda$ tends pointwise to $1$. 
Limited resolution of figure drawing misses the fact 
that $f_\lambda(0,\kappa) = f_\lambda(1,\kappa) = 0$ for 
$m > 0$ (see the text).}
\end{center}
\end{figure}
the factor 3 in Eq. (\ref{HlambdaI5}). This means
that the factor 3 appears in front of one and the 
same operator that is obtained in turn from each 
of the three regions equally. Appendix \ref{a:connection} 
describes changes of variables that exhibit the equivalence 
of these three regions of integration over $\zeta$. 

Fig. 1 
% \ref{fig:FZetaKappa} 
shows the function
$f_\lambda \equiv f_\lambda(\zeta,\kappa^\perp) =
\exp{ [ - (\Delta {\cal M}^2/\lambda^2 )^2 ] }$,
where $\Delta {\cal M}^2$ as function of $\zeta$
and $\kappa^\perp$ is given in Eq.
(\ref{DeltaMain}), for six different values of
$\lambda$: 50, 10, 4, 3, 2, and 1.5, in units of
the mass $m$. Momentum $\kappa^\perp$ is also
given in units of $m$. The six three-dimensional plots
are approximate because of a limited resolution of
figure drawing. In particular, $f_\lambda = 0$ for
$\zeta=0$ and $\zeta=1$ for all values of
$\lambda$, except that the region where
$f_\lambda$ is very small decreases in size and
relevance in the Fourier integral when $\lambda
\rightarrow \infty$. This means that the
non-locality of the effective interaction
Hamiltonian $H_{\lambda I}$ can be systematically
considered only under the assumption that the
domain of $H_{\lambda I}$ does not contain states
with wave functions that are significantly
singular as functions of $\zeta$ at 0 and 1. We
assume that the domain of $H_{\lambda I}$ obeys
this condition, on the basis of our expectation
that a non-zero mass-term leads to a spectrum that
satisfies this condition.

The middle bumps on the six plots in Fig. 1 
% \ref{fig:FZetaKappa} 
show the RGPEP form factor $f_\lambda$ in 
the range $0 < \zeta < 1$. In this range, $\zeta$ 
is the same as the parameter $x$ typically used 
in LF notation for $+\,$-momentum fractions. The 
other two bumps that are visible in all six plots 
in the regions $\zeta < 0$ and $\zeta > 1$, combine 
together with the bump in the middle to form a 
function that approaches 1 pointwise when $\lambda 
\rightarrow \infty$. This is how one obtains a 
local interaction in this limit. 

It is visible in Fig. 1 % \ref{fig:FZetaKappa}
that for $\lambda \rightarrow \infty$ one obtains
interaction whose strength approaches a constant 
in momentum space, and hence tends to a point-like 
interaction in position space. When the position 
probe one uses has a resolution considerably below
$1/\lambda$, the non-locality corresponding to 
$\lambda$ remains invisible. The greater $\lambda$, 
the smaller distance scales at which the interaction 
still appears to be local, i.e., the interaction 
density spreads only over regions that are smaller 
than the probe can resolve. 

It is also visible in Fig. 1
%\ref{fig:FZetaKappa}  
that the smaller $\lambda$ the greater overall 
suppression of the strength of the interaction 
by the form factor $f_\lambda$. This suppression 
can be reinforced (triviality) or reduced 
(asymptotic freedom) by the variation of the 
coupling constant $g_\lambda$ when $\lambda$ 
decreases.

%%%%%%%%%%%%%%%%%%%%%%%%%%%%%%%%%%%%%%%%%%%%%%%%%%%%%%%%%%%%
%%%%%%%%%%%%%%%%%%%%%%%  SUBSECTION  %%%%%%%%%%%%%%%%%%%%%%%
%%%%%%%%%%%%%%%%%%%%%%%%%%%%%%%%%%%%%%%%%%%%%%%%%%%%%%%%%%%%

\subsection{ Non-locality for $m \rightarrow 0$ }
\label{m->0}

High-energy dynamics involves interactions in
which spatial momenta of particles are typically
very large in comparison with their masses. In
such circumstances, it may be useful to neglect
the masses. One can also consider masses that are
negligible in comparison with spatial momenta of
interacting particles, even if the latter are not
very large on the scale of momenta measurable in
laboratory. This section discusses non-locality of
first-order RGPEP three-particle vertices in the
case when masses are neglected.

The case of $m=0$ can be seen as resulting from
the limit $m \rightarrow 0$ in the sense that the
range of relative momenta accessible to particles
created or annihilated in a single act of
interaction is much greater in size than $m$.
Since the invariant mass depends on $q^{\perp 2} +
m^2$, large momentum $q^\perp$ makes a small mass
parameter $m$ irrelevant for the value of the form
factor $f_\lambda$. But for small momentum
$q^\perp$, the mass $m$ may be important because
$m^2$ is divided by $z$ or $1-z$ in the invariant
mass. For $q^\perp=0$, this means that an
arbitrarily small $m$ matters in the value of
$f_\lambda$ wherever $m^2/z$ or $m^2/(1-z)$ are
not negligible in comparison with $\lambda$. For
$m \ll \lambda$, this happens only for extreme
values of $z$. Therefore, one can expect that the
limit $m \rightarrow 0$ is equivalent to the
result of setting $m=0$ with the exception of
contributions from the end points in $z$ while
$|q^\perp| \lesssim m$. At the end points, the
case $m=0$ qualitatively differs from all cases
with $m > 0$. The former case allows for non-zero
contributions from the end points when $q^\perp
\rightarrow 0$, and all the other cases do not. 

One can gain some understanding of the
non-locality corresponding to $\lambda \gg m$ by
neglecting $m$ entirely and replacing the form
factor $f_\lambda = \exp[-({\cal M}_{12}
/\lambda)^4]$ by $f_\lambda = \exp[-({\cal
M}_{12}/\lambda)^2]$ in Eq. (\ref{barf}). Such
replacement preserves a non-local nature of the
vertex qualitatively and greatly simplifies
calculations. The simplified case is discussed in
this section.

Under the simplifying assumptions, the non-local 
vertex function in Eq. (\ref{barf}) reads
\beq
\label{fm0}
\bar f_\lambda(x_1, x_2 , x_3)
\es
3
\int { d^3 P \, \theta(P^+) \over 2 (2\pi)^3} 
\int_0^1 {dz \, P^+         \over 2 (2\pi)  } 
\int {d^2q^\perp            \over   (2\pi)^2} 
\,
e^{- {q^{\perp 2} \over \lambda^2 z (1-z)} } 
\nt
e^{-iP[ z x_1 + (1-z) x_2 - x_3] + i q^\perp (x_1^\perp - x_2^\perp)} \, .
\eeq
Eq. (\ref{a:fm0}) in Appendix \ref{a:m=0} provides
the result of integration over momenta $q^\perp$, 
$P^\perp$, and $P^+$.\footnote{ Integration over $P^+$ 
requires a regularization that results in the presence
of $i\epsilon$ in Eq. (\ref{explicitgrt}); see Appendix 
\ref{a:m=0}. Despite the presence of $i\epsilon$, the 
resulting interaction Hamiltonian is Hermitian
because Eq. (\ref{HlambdaI6}) contains a term with 
$\bar f_\lambda$ and a conjugated term, denoted by
$h.c.$} The vertex function is invariant under 
translations and depends on 6 relative position 
co-ordinates. For example, when one identifies a 
point of reference with the argument of the field 
labeled 2, the result takes the form
\beq
\label{grt}
\bar f_\lambda(x_1, x_2 , x_3)
\es
\lambda^6 \, g[\lambda(x_1 - x_2), \lambda(x_3 - x_2)] \, ,
\eeq
where
\beq
\label{explicitgrt}
g(\rho, \tau) 
\es
{ - 3 \over 16 \pi^3 } 
\int_0^1 
{dz \, z(1-z) 
\over 
( \tau^- - z \rho^- + i\epsilon )^2 } 
\,
\delta^2(\tau^\perp  - z \rho^\perp)
\,
e^{- z(1-z)\,\rho^{\perp 2}/4} \hspace{-1mm}.
\eeq
The function $g(\rho, \tau)$ is invariant with 
respect to rotations around $z$-axis. It is 
different from 0 only for $\tau^\perp$ being 
a fraction of $\rho^\perp$. 

In order to visualize the non-local vertex
function $\bar f_\lambda(x_1, x_2 , x_3)$, it is
sufficient to consider it a function of $\tau$ 
on the LF hyper-plane of $\rho$. Moreover, since 
$\tau^\perp = t \rho^\perp$ with $t = |\tau^\perp|
/|\rho^\perp|$ in the range $0 \leq t \leq 1$, 
one can visualize $\bar f_\lambda(x_1, x_2 , x_3)$
by drawing it on the two-dimensional plane of 
only two variables: $\tau^\perp || \rho^\perp$ 
and $\tau^-$. Namely, using the decomposition 
\beq
\tau^\perp \es t \, \rho^\perp + s \, n^\perp  \, ,
\eeq
\begin{figure}
\label{fig:DistanceMassLess}
\begin{center}
 \epsfig{file=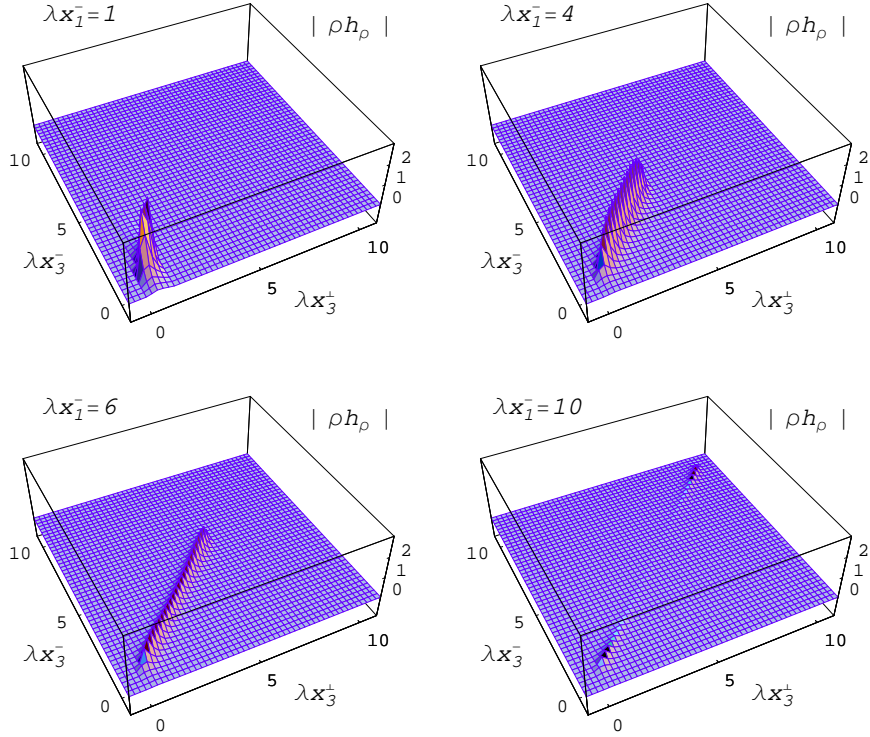,scale=1.2}
\caption{ 
Modulus of the non-local interaction density 
for massless particles, $\bar f_\lambda (x_1, x_2, 
x_3)$ in Eq. (\ref{fm0}), drawn as function of 
$\lambda x_3^\perp$ and $\lambda x_3^-$ on the 
space-time plane containing $x_1^\perp - x_2^\perp$ 
in the LF hyper-plane defined by the condition 
$x^+=0$, in terms of the function $h_\rho$ in Eq. 
(\ref{hrho}) multiplied by $|\rho^\perp|$ for
visualization of the end-point enhancement, for 
$\lambda x_2 = 0$ and several choices of the 
points $\lambda x_1$ with $\lambda x_1^- = 
\lambda |x_1^\perp| = 1$, 4, 6, and 10, as indicated. 
The end-point enhancement for massless particles
when $x_3$ approaches $x_1$ or $x_2$ shows up when 
the dimensionless distance $\lambda|x_1^\perp - 
x_2^\perp|$ exceeds 4. For further explanation, 
see Eqs. (\ref{grt}) to (\ref{t}) and the text.}
\end{center}
\end{figure}
with $n^\perp \rho^\perp = 0$ and $n^\perp n^\perp = 1$, 
one can write
\beq
\label{explicitgrt1}
g(\rho, \tau) 
\es
\delta(s) \,\, 
{ 3 \over 16\pi^3 } \,\,
h_\rho (t, \tau^-) \, ,
\eeq
where
\beq
\label{hrho}
h_\rho(t, \tau^-)
\es
{ - t(1-t) 
\over 
( \tau^- - t \rho^- + i\epsilon )^2 } 
\,
{ e^{- t(1-t)\,\rho^{\perp 2}/4} 
\over |\rho^\perp| } \, , \\
\label{t}
0 \leq t \es |\tau^\perp|/|\rho^\perp| \leq 1 \, .
\eeq
For every choice of $\rho = \lambda (x_1-x_2) = 
(\rho^\perp , \rho^-)$, the function $h_\rho
( t, \tau^-)$ is a function of $ 0 < t < 1 $ 
and $t^-$. Figs. 2
% \ref{fig:DistanceMassLess} 
and 
% \ref{fig:AngleMassLess} 
3 show examples
that illustrate generic features of 
$\bar f_\lambda (x_1, x_2, x_3)$.

Fig. 2 
% \ref{fig:DistanceMassLess} 
illustrates
how the non-local interaction strength on the LF
depends on the position of the point $x_3$, where
particle number 3 is annihilated (created), for a
given distance between the two points $x_1$ and
$x_2$, where particles 1 and 2 are created
(annihilated). The distances are measured in
dimensionless units that result from
multiplication of position coordinates by the
RGPEP scale parameter $\lambda$. To obtain the
Hamiltonian corresponding to $\lambda$, the
density $h_\rho$ whose modulus times $|\rho^\perp|$
is shown in Fig. 2,
% \ref{fig:DistanceMassLess} 
multiplied by the
factors present in Eq. (\ref{explicitgrt1}) to
obtain $\bar f_\lambda (x_1, x_2, x_3)$, is
integrated with the product of three effective
fields $\psi_\lambda(x)$ using $\int \lambda^3
d^3x_1 \int \lambda^3 d^3x_2 \int d^3x_3$, see Eq.
(\ref{HlambdaI6}). 

For the purpose of drawing, Fig. 2
% \ref{fig:DistanceMassLess} 
is artificially modified in so far that
the infinitesimal parameter $\epsilon \rightarrow
0_+$ is set to 1/5. This substitution causes
that the rise of the modulus of $\bar f_\lambda
(x_1, x_2, x_3)$ due to the square of $x_3^- -
x_2^- - (x_1^- - x_2^-)|x_3^\perp - x_2^\perp|/
|x_1^\perp - x_2^\perp|$ crossing 0 in denominator
is limited to $1/\epsilon^2 = 25$, instead of
reaching $0_+^{-2}$. The artificial modification
preserves generic features of $\bar f_\lambda
(x_1, x_2, x_3)$. Namely: a) $\bar f_\lambda (x_1,
x_2, x_3)$ pointwise decreases with distance 
between $x_1^\perp$ and $x_2^\perp$, b) it quickly
vanishes outside the region where $x_3$ is near a
straight line connecting $x_1$ with $x_2$ on the
LF, and c) it is notably spread toward the end-points 
when $\lambda |x_1^\perp - x_2^\perp|$ exceeds 4. 
The number 4 results from the factor $e^{-
t(1-t)\,\rho^{\perp 2}/4}$ in Eq. (\ref{hrho}).
This factor is varying slowly until
$|\rho^\perp|^2$ exceeds the inverse of maximal
value of $t(1-t)/4$, which is 16. Thus, when
$|\rho^\perp|$ exceeds 4, the density $\bar
f_\lambda (x_1, x_2, x_3)$ begins to be suppressed
in the middle between $x_1$ and $x_2$. When
$|\rho^\perp|$ increases far above 4, the
interaction density favors $x_3$ near $x_1$ or
$x_2$, being squeezed in the middle between 
$x_1$ and $x_2$.

Fig. 2
% \ref{fig:DistanceMassLess} 
illustrates the feature that $x_3$ must lie near 
a line connecting $x_1$ and $x_2$ on the LF in 
the cases where $|x_1^- - x_2^-| \sim |x_1^\perp 
- x_2^\perp|$. 
Fig. 3
% \ref{fig:AngleMassLess} 
shows that the same happens in other cases, by 
providing examples for different orientations 
of $x_1 - x_2$ on the LF. Again, $\epsilon$ is set 
to $1/5$ in order to avoid infinite values on the
line connecting points $x_1$ and $x_2$, which is 
the same trick of convenience in drawing that 
was used earlier in Fig. 2
% \ref{fig:DistanceMassLess} 
and is further explained below.
\begin{figure}
\label{fig:AngleMassLess}
\begin{center}
 \epsfig{file=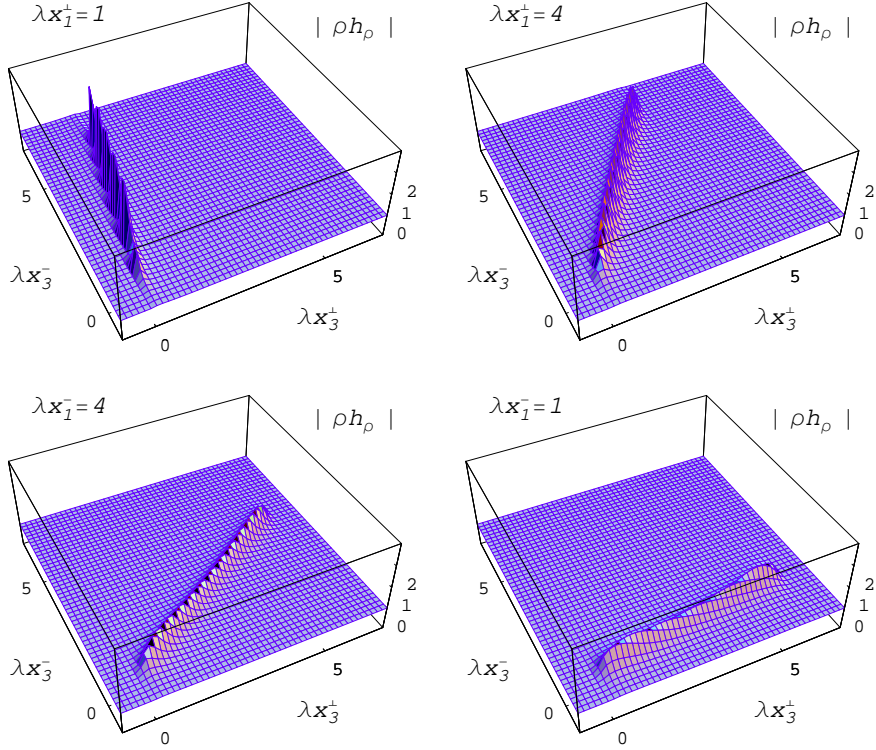,scale=1.2}
\caption{ 
The same non-local interaction density function 
as in Fig. 2,
% \ref{fig:DistanceMassLess} 
also multiplied by $|\rho^\perp|$ 
as in Fig. 2 to avoid suppression on the drawing 
when the distance between $x_1^\perp$ and $x_2^\perp$ 
increases, plotted for several choices of the points 
$\lambda x_1$ with 
$\lambda (|x_1^\perp|, x^-) = 
 (1,6)$,
$(4,6)$,
$(6,4)$, and
$(6,1)$, as indicated. 
See the text and Eqs. (\ref{grt}) to (\ref{t}).}
\end{center}
\end{figure}

Both Figs. 2
% \ref{fig:DistanceMassLess} 
and 3
% \ref{fig:AngleMassLess} 
display partly jagged shapes of the drawn
functions, which requires explanation. For 
this purpose, consider the function
\beq
\label{fden0}
f(x) \es -(x + i\epsilon)^{-2} \, ,
\eeq
in the vicinity of $x=0$. Fig. 4
% \ref{fig:fZero}
displays the modulus and real part of this 
function for $\epsilon = 1/5$.
\begin{figure}
\label{fig:fZero}
\begin{center}
 \epsfig{file=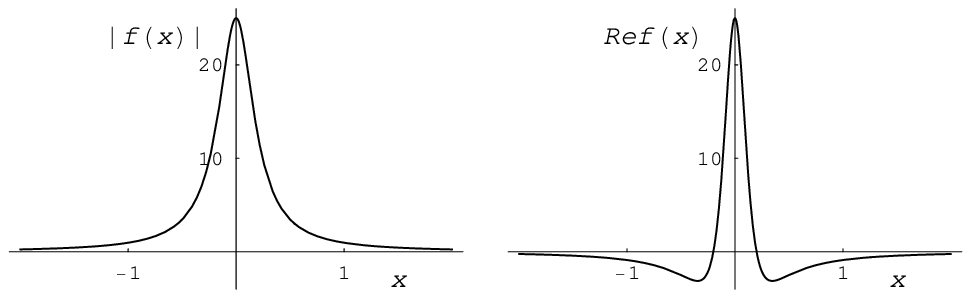,scale=1.2}
\caption{ 
Modulus and real part of the function $f(x)$
in Eq. (\ref{fden0}). Shape of $|f(x)|$ explains 
the somewhat jagged shapes of functions
displayed in Figs. 2
% \ref{fig:DistanceMassLess} 
and 3
% \ref{fig:AngleMassLess}
; see the text.}
\end{center}
\end{figure}
Since Figs. 2
% \ref{fig:DistanceMassLess} 
and 3
% \ref{fig:AngleMassLess} 
are drawn using a crude, uniform mesh of points 
for arguments and a uniform trapezoidal interpolation 
for shading of the function image, it happens from 
time to time that the mesh selects an argument point 
a bit away from a narrow peak (forming a wall of a 
priori infinite height) in the plotted function. The 
selected argument produces a small value of the 
function. When such small value is crudely connected 
with large values at neighboring points, a somewhat 
jagged shape is obtained. When the mesh spacing is 
decreased, drawings become smooth. However, the files 
with a small mesh spacing have a prohibitive size. 
Figs. 1,
%\ref{fig:FZetaKappa} 
2,
% \ref{fig:DistanceMassLess} 
and 3
% \ref{fig:AngleMassLess}  
are provided as a result of a compromise between
precision of rendering and size of the figure files.
The compromise is made in such a way that in every
case the key feature to be displayed and discussed
is not altered in any significant way. 

%%%%%%%%%%%%%%%%%%%%%%%%%%%%%%%%%%%%%%%%%%%%%%%%%%%%%%%%%%%%
%%%%%%%%%%%%%%%%%%%%%%%  SUBSECTION  %%%%%%%%%%%%%%%%%%%%%%%
%%%%%%%%%%%%%%%%%%%%%%%%%%%%%%%%%%%%%%%%%%%%%%%%%%%%%%%%%%%%

\subsection{ Non-locality for $\lambda \sim m$ }
\label{lambda<m}

In attempts to understand mass generation in a
relativistic theory of particles, such as attempts
to solve QCD for quark and gluon wave functions of
hadrons in the Minkowski space or attempts to
resolve the strong interaction structure of
photons that results in mixing of photons with
$\rho$-mesons, one is faced with a challenge of
understanding dynamics of binding of effective
constituents of some mass $m$. In QCD, the
effective constituent mass $m$ that dominates the
mechanism of binding of lightest quarks appears to
be of the order of $\Lambda_{QCD}$. The problem of
chiral symmetry breaking in QCD and other theories
can be rephrased as a question of what are the
dominant interaction terms in an effective
Hamiltonian of QCD when the decreasing RGPEP
parameter $\lambda$ becomes comparable with
$\Lambda_{QCD}$. In any case, unless the parameter
$\lambda$ is comparable with or smaller than the
constituent mass $m$, a large number of
constituents may participate in the dynamics
described by Hamiltonians evaluated using RGPEP.
Therefore, it is of interest to construct and
understand the structure of Hamiltonian
interaction terms with small $\lambda$, where by
small it is meant that $\lambda$ is comparable
with $m$. This section discusses the non-locality
of first-order RGPEP three-particle vertex in this
regime.

It is visible in Fig. 1
% \ref{fig:FZetaKappa} 
that the RGPEP momentum-space form factor $f_\lambda$ 
peaks at $\zeta \sim 1/2$, i.e., in the center of a 
middle bump in Fig. 1.
% \ref{fig:FZetaKappa} 
Section \ref{How} shows that the two neighboring
regions with $\zeta$ below 0 and above 1 both
contribute the same operator that the region $0 <
\zeta < 1$ contributes. As $\lambda$ decreases,
the size of $f_\lambda$ also decreases. The
smaller $f_\lambda$ the weaker the interaction
that changes the number of effective particles.
However, in asymptotically free theories, the
effective coupling constant $g_\lambda$ increases
when $\lambda$ decreases; $g_\lambda$ may partly
compensate for the small size of $f_\lambda$ at
its peak. The resulting interaction strength may
thus be not eliminated entirely when $\lambda$
decreases to about $m$. Instead, the strength is
located in the domain in momentum space where
$f_\lambda$ is maximal. Therefore, for $\lambda
\sim m$ or smaller, given that the three regions
of $\zeta < 0$, $0 < \zeta < 1$, and $1 < \zeta$
contribute the same operator to the Hamiltonian,
it is sufficient to consider the middle region of
$\zeta$ in Fig. 1, 
% \ref{fig:FZetaKappa}
i.e., the region where $f_\lambda$ forms a bump 
around $z \equiv \zeta \sim 1/2$. In this region, 
the relative transverse momentum $q^\perp \equiv 
\kappa^\perp$ is limited to values not exceeding 
order $\lambda$, tempered in addition by how much 
the variable $z$ deviates from 1/2.

For a closer inspection of the middle bump region
for $\lambda \sim m$, consider the form factor
\beq
f_\lambda
\es
e^{ - \left[ {\cal M}_{12}^2 - m^2
\right]^2 / \lambda^4 } \, ,
\eeq
with ${\cal M}_{12}$ given in Eq. (\ref{M12}), 
rewritten as 
\beq
f_\lambda \es e^{-(3m^2/\lambda^2)^2} \,
e^{ - \left[ ({\cal M}_{12}^2 - m^2 )^2 
           - (3m^2)^2 \right]/\lambda^4  } \, .
\eeq
The exponential in front describes the size of the form 
factor at its maximum at $z = 1/2$ and $q^\perp = 0$,
while the remaining factor describes the form factor 
fall-off away from its maximum. Now,
\beq
({\cal M}_{12}^2 - m^2 ) ^2 -(3m^2)^2 
\es
({\cal M}_{12}^2  + 2m^2)
({\cal M}_{12}^2  - 4m^2) \, , 
\eeq
where the first factor on the right-hand side is not 
smaller than $6m^2$ while the second factor can be 
small. These two factors can be analyzed in detail 
in terms of variables 
\beq
\label{kp}
k^\perp & = & q^\perp \, , \\
\label{kz}
k^z     & = & (z-1/2) \, {\cal M}_{12} \, , \\
\label{dz}
dz \es 4z(1-z) {dk^z \over {\cal M}_{12} } \, ,
\eeq
with which
\beq
\label{zk}
z \es {1 \over 2} \left( 1 + {k^z \over \sqrt{m^2 +
\vec k^2} } \right) \, , \\
{\cal M}^2_{12}  \es 4 ( \vec k \, ^2 + m^2 ) \, ,
\eeq
and 
\beq
f_\lambda \es e^{-(3m^2/\lambda^2)^2} \,
e^{ - { \vec k\,^2 \, + \,\, 3m^2/2 \over (\lambda/2)^2 } \, {
\vec k\,^2 \over (\lambda/2)^2 } } \, .
\eeq
For $\lambda \sim m$ or smaller, the relative momentum 
$|\vec k\,|$ of particles 1 and 2 is smaller than 
$\lambda/2 \sim m/2$ and the RGPEP form factor can 
be very well approximated by Gaussian
\beq
f_\lambda \es e^{-9m^4/\lambda^4} \, e^{ - 24 m^2 \vec k\,^2 /\lambda^4 }  \, .
\eeq
Using this approximation in Eq. (\ref{barf}) written 
in terms of variables $R$ and $r$ defined in Eqs. 
(\ref{R}) to (\ref{x2}), one obtains
\beq
\bar f_\lambda(x_1, x_2, x_3)
\es
3\,
e^{- 9 m^4/\lambda^4 } \,
\int {d^3P \, P^+ \theta(P^+) \over 2 (2\pi)^3} \, 
 e^{ - i P(R-x_3) } 
\nt
\int { 4z(1-z) \, d^3k
\over 2 (2\pi)^3 {\cal M}_{12}}
\,
e^{ - 24 m^2 \vec k\,^2 /\lambda^4 } 
\,\,
e^{ i [-(z-1/2)Pr + k^\perp r^\perp] } \, .
\eeq
Since for $\lambda \sim m$ or smaller one 
has $z \sim 1/2$ and
\beq
k^z \es (z-1/2) {\cal M}_{12} \sim (z-1/2) 2m \, ,
\eeq
the function  $\bar f_\lambda(x_1,
x_2, x_3)$ is approximated by 
\beq
\bar f
\es
3 \, 
e^{- 9m^4/\lambda^4} \,
\int {d^3P P^+ \theta(P^+) \over 2 (2\pi)^3} \, 
e^{ - i P(R-x_3) } 
\nt
\int { d^3k 
\over 4m (2\pi)^3 }
\,
e^{ - 24 m^2 \vec k\,^2 /\lambda^4 } 
\,\,
e^{ i [k^z (-Pr/(2m)) + k^\perp r^\perp] } \, .
\eeq
The key observation is that one can introduce a three-vector 
\beq 
\label{threevectorr}
\vec r \es \vec r \, (P, r) \rs \left( {- Pr \over 2m}, r^\perp \right) \, ,
\eeq
and write
\beq
\label{fr}
\bar f
\es
3 \, e^{- 9 m^4/\lambda^4 } \,
\int {d^3P P^+ \theta(P^+) \over 2 (2\pi)^3} \, 
e^{ - i P(R-x_3) } 
\nt
\int { d^3k 
\over 4m (2\pi)^3 }
\,
e^{ - 24 m^2 \vec k\,^2 /\lambda^4 } 
\,\,
e^{ i \vec k \, \vec r } \, ,
\eeq
so that the non-locality is parameterized in terms of $\vec r$.
Details of evaluation of the non-locality are described in 
Appendix \ref{a:lambda=m}. Like in Eqs. (\ref{grt}) to 
(\ref{t}) for massless particles, one can write
\beq
\label{fmassive}
\bar f_\lambda(x_1, x_2, x_3) 
= \lambda^6 \, g(\rho,\tau) \, ,
\eeq
\begin{figure}
\label{fig:DistanceMassive}
\begin{center}
 \epsfig{file=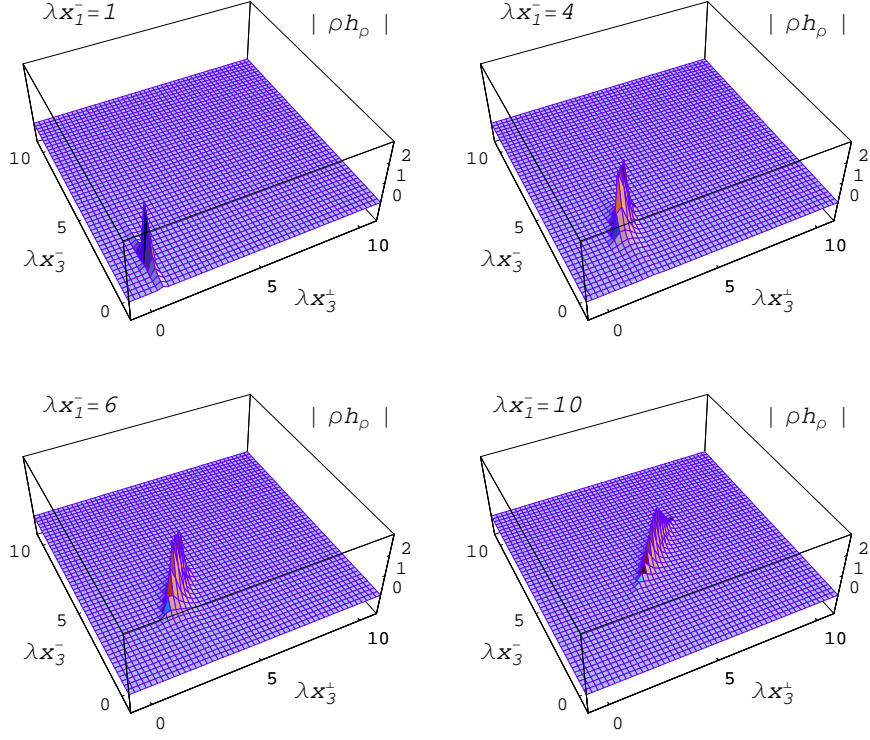,scale=1.2}
\caption{ 
Modulus of the non-local LF interaction density 
$\bar f_\lambda (x_1, x_2, x_3)$ in Eq. (\ref{fmassive})
for massive particles with $\lambda = m$, for which 
$\gamma = 1/6$, using the same convention as in Fig. 2
% \ref{fig:DistanceMassless}
but in terms of $|\rho^\perp|$ times $h_\rho$ in Eq. 
(\ref{hm}) multiplied in addition by $e^{1\over 4 
\gamma^2}/\gamma$. The extra multiplication is needed 
for removing the exponential suppression of the form 
factor $h_\rho$ so that the height of the plotted 
function matches with the one in Fig. 2. 
% \ref{fig:DistanceMassless}
For further explanation, see Eqs. (\ref{fmassive}) 
to (\ref{hm}) and the text.}
\end{center}
\end{figure}
where $\tau^\perp = t \rho^\perp + s n^\perp$ and
\beq
g(\rho,\tau)
\es
\delta (s)
\,
{3 \over 16 \pi^3} 
\, 
h_\rho(t, \tau^-) \, .
\eeq
Using $\gamma = {\lambda^2  \over 6 m^2}$, Eq. (\ref{ahm})
implies
\beq
\label{hm}
h_\rho(t, \tau^-)
\es
{ - {1 \over 4} \, e^{- { 8 \over 3 \gamma^2} \, \left( t - 1/2 \right)^2  } 
\over 
\left( \tau^- - t \rho^-  + i \epsilon \right)^2 }
\,
{ e^{- \gamma \rho^{\perp 2}/16 }
\over |\rho^\perp| }
\,\,
{ \gamma \over e^{1\over 4 \gamma^2 } }
\, . 
\eeq
For $\gamma =1$ (or $\lambda = \sqrt{6} \, m$), 
this result matches the massless case of Eq.
(\ref{hrho}) up to the factor $e^{-1/4}$ when 
$t \sim 1/2$. In this case, $t(1-t)$ amounts to 
1/4. 

Non-locality of the interaction vertex for 
$\lambda = m$ is illustrated in Fig. 5,
%\ref{fig:DistanceMassive}
which should be compared with Fig. 2.
% \ref{fig:DistanceMassLess} 
For $\lambda x_3$ away from the points with 
$t \sim 1/2$, especially at the end-points, the 
result for $\lambda = m$ shown in Fig. 5 
% \ref{fig:DistanceMassive} 
is quite different from the massless case
shown in Fig. 2; 
% \ref{fig:DistanceMassLess} 
there is a strong suppression instead of 
enhancement at the end points. The smaller 
$\gamma$, i.e., the smaller $\lambda/m$, the 
narrower the peak at $t=1/2$ and lesser 
exponential suppression of the non-local 
vertex due to increase of $|\rho^\perp|$. 

In contrast to Fig. 2,
% \ref{fig:DistanceMassLess} 
Fig. 5
% \ref{fig:DistanceMassive}
exhibits the feature that the non-local 
interaction density prefers the position 
of $x_3$ in the middle of $x_1$ and $x_2$.
This is a characteristic behavior for 
non-relativistic interactions, i.e., 
interactions in which relative momentum
of interacting particles is smaller than 
their masses. Interactions of relativistic
particles do not have this feature.

Finally, one can observe that the 
three points $x_1$, $x_2$, and $x_3$ 
must approximately lie on a straight line
on the LF. This feature is the same as in 
Fig. 3
% \ref{fig:AngleMassLess} 
and does not require a separate drawing.

%%%%%%%%%%%%%%%%%%%%%%%%%%%%%%%%%%%%%%%%%%%%%%%%%%%%%%%%%%%%
%%%%%%%%%%%%%%%%%%%%%%%  SUBSECTION  %%%%%%%%%%%%%%%%%%%%%%%
%%%%%%%%%%%%%%%%%%%%%%%%%%%%%%%%%%%%%%%%%%%%%%%%%%%%%%%%%%%%

\subsection{ Non-locality in a non-relativistic theory }
\label{nr}

There is an analogy with a non-relativistic
quantum mechanics that may be helpful in
interpretation of the results described in the
previous sections. The difficulty of
interpretation may be expected because RGPEP is
developed in the front form of Hamiltonian
dynamics and the non-locality of interaction
Hamiltonian densities is obtained on the LF
hyper-plane in space-time rather than in space at
a single moment of time. Thus, the relativistic
RGPEP involves concepts that are general enough
for hoping that the method will apply in
derivation of wave functions of bound states of
partons in theories with asymptotic freedom and
infrared slavery or in attempts to understand
symmetry breaking and mass generation at high
energies. The space-time concepts do not appear
quite intuitive from the point of view based on
non-relativistic quantum mechanics. The latter
only applies when motion of charged particles is
characterized by velocities $v \sim \alpha_{QED}
c$, where $c$ is the speed of light and
$\alpha_{QED}$ is the fine structure constant
$\sim 1/137$. This constant determines the
strength of interactions which govern behavior of
electrons bound in atoms. Binding of quarks and
gluons occurs in space-time as a result of
interactions with coupling constants about 100
times greater than in QED and the non-relativistic
intuition is not directly applicable. The same
difficulty with building a physical picture is
faced in all theories of mass generation and
related symmetry breaking.

Fortunately, an intuitive picture to think about
can be arrived at by comparing RGPEP form factors
$f_\lambda$ for small $\lambda$ with simple form
factors that may be introduced ad hoc in an
effective non-relativistic theory. For example,
consider interactions that resemble emission and
absorption of mesons by nucleons in nuclear
physics. Suppose an effective interaction
Hamiltonian has the form 
\beq
\label{Hnr}
H_{\lambda I} \es 
g_\lambda \,
\left[ \prod_{i=1}^3 \int {d^3 p_i \over (2\pi)^3 } \right] 
(2\pi)^3 \, \delta^3(p_1 + p_2 - p_3) 
\nt  
f_\lambda \,\, (a^\dagger_{\lambda p_1} \, 
               a^\dagger_{\lambda p_2} \, 
               a_{\lambda p_3} + h.c. )\, , 
\eeq
where $\vec p_i, i =1$, 2, 3 are standard 
three-dimensional momentum variables conjugated
with standard rectilinear co-ordinates 
$\vec x_i, i =1$, 2, 3 in space, respectively.
Let an arbitrarily chosen form factor be
\beq
f_\lambda
\es
e^{ - (\vec p_1 - \vec p_2  )^2 / \lambda^2 } \, .
\eeq
Suppose that one builds effective fields at the 
moment $t=0$ using operators $a_\lambda$ (only 
annihilation operators) and evaluates a non-local 
Hamiltonian density in space. Instead of Eq. 
(\ref{barf}), one is led to consider an expression 
of the form 
\beq
\bar f_\lambda(x_1, x_2, x_3)
& \sim &
  \int {d^3 P \over (2\pi)^3 } 
  \int {d^3 q \over (2\pi)^3 } 
\, f_\lambda \, e^{ - i X } \, , 
\nn
\\
X \es (P/2+q) x_1 + (P/2-q)x_2 - P x_3 \\
  \es P[(x_1+x_2)/2 -x_3) + q(x_1-x_2) \, ,
\eeq
and obtains
\beq
\bar f_\lambda(x_1, x_2, x_3)
& \sim &
\delta^3[\vec x_3 - (\vec x_1+\vec x_2)/2] \,
\,
\lambda^3 \, e^{ - \lambda^2 (\vec x_1-\vec x_2)^2/16 } \, .
\eeq

This result has an intuitive interpretation.
Particle 3 is annihilated (created) exactly in 
the middle of positions $x_1$ and $x_2$ where
particles 1 and 2 are created (annihilated),
respectively. The width of the distribution of 
distances between points $x_1$ and $x_2$ is 
$1/\lambda$. When $\lambda \rightarrow \infty$, 
the interaction is local. When $\lambda$ becomes
small, it is greatly delocalized in the sense
that the distance between $x_1$ and $x_2$ can
be large, but at the same time $x_3$ is always
precisely in the middle between $x_1$ and $x_2$.

The situation is somewhat similar to the one shown
in Fig. 5, 
% \ref{fig:DistanceMassive} 
except that in Fig. 5 one of the directions is
$x_3^-$ on the LF. However, when $\lambda$ is
small in comparison with masses and $x_3$ is
fairly located on the LF between $x_1$ and $x_2$,
one can observe that for all co-ordinates involved
$x^+=0$ and $x^- = - 2x^3$. Therefore, in the
non-relativistic case, $r^-$ can be seen as
analogous to $-2r_z$. This way one recovers the 
simple non-relativistic interpretation of the LF
non-locality. Further discussion is provided in
the next section.

%%%%%%%%%%%%%%%%%%%%%%%%%%%%%%%%%%%%%%%%%%%%%%%%%%%%%%%%%%%%
%%%%%%%%%%%%%%%%%%%%%%%  SUBSECTION  %%%%%%%%%%%%%%%%%%%%%%%
%%%%%%%%%%%%%%%%%%%%%%%%%%%%%%%%%%%%%%%%%%%%%%%%%%%%%%%%%%%%

\subsection{ Relationship to a 2-body wave function }
\label{wavefunction}

Previous section provided an interpretation of
the non-local interaction Hamiltonian densities 
on the LF by analogy that was limited to slow 
particles. This section provides another intuitive 
picture that is more precise and not limited to 
slowly moving bound states. In fact, it can be 
used in any frame one wishes to use, including 
the infinite momentum frame (IMF) and the center 
of mass frame (CMF) alike.

In order to relate the non-locality of an
effective Hamiltonian interaction term to a wave
function of a 2-body bound state on the LF
hyperplane defined by condition $x^+ = 0$, we
introduce three species of particles which are
annihilated by operators $a_1$, $a_2$, and $a_3$,
respectively. The three species are introduced to
avoid the need for symmetrization of relevant
functions for identical particles. Masses of the
three species are assumed all equal $m$ in order
to match the conditions set in the previous
sections.

Consider the matrix element 
\beq
\label{vf}
\phi_{\lambda P}(x_1, x_2) \es 
\langle 0 | \psi^{+}_{1\lambda}(x_1) \, \psi^{+}_{2\lambda}(x_2) 
\, H_{\lambda I} \, a^\dagger_{3\lambda P}|0 \rangle \, , 
\eeq
where $|0\rangle$ denotes the vacuum state that is
annihilated by $a_{1\lambda p}$, $a_{2\lambda p}$,
and $a_{3\lambda p}$ for $p^+>0$. The effective 
quantum fields $\psi_{1\lambda}(x)$, $\psi_{2\lambda}(x)$, 
and $\psi_{3\lambda}(x)$ are defined by Eqs.
(\ref{psil}) to (\ref{alp}) for species 1, 2, and 3,
respectively. The superscript $+$ means differentiation,
\beq
\psi^{+}(x) \es i\partial^+ \psi(x) \, .
\eeq
The state $|P\rangle$ with $P^+>0$, defined 
as a particle of the third species 
\beq
|P\rangle \es a^\dagger_{3\lambda P} |0\rangle \, ,
\eeq
can be considered analogous to a bound state of two
effective constituents of species 1 and 2 corresponding 
to scale $\lambda$. Namely, the 
eigenvalue equation for a bound state with momentum $P$, 
\beq
\left( H_{\lambda 0} + H_{\lambda I} \right)|P \rangle 
\es P^- |P \rangle \, ,
\eeq
can be rewritten as 
\beq
|P\rangle \es {1 \over P^- - H_{\lambda 0} } \,  
H_{\lambda I} |P\rangle \, .
\eeq
The Hamiltonian $H_{\lambda I}$ couples 
a pair of particles 1 and 2 with a particle
3. Therefore, the eigenvalue equation for the state
$|P\rangle$ involves both the 2-particle component 
of type 1 and 2 and the 1-particle component
of type 3. This LF situation resembles the Lee model
\cite{LeeModel,4thorderRGPEP}. The wave function to 
focus on is the 2-particle component. One can insert 
the identity in 2-body space,
\beq
1_{12} 
\es 
\int d^3x_1 \, d^3 x_2 \,\, 
\psi_{1\lambda}(x_1) \, \psi_{2\lambda}(x_2) |0\rangle  
\langle 0| \psi^{+}_{1\lambda}(x_1) \, \psi^{+}_{2\lambda}(x_2)
\, ,
\eeq
on the left-hand side of $H_{\lambda I}$ and evaluate the 
matrix element 
\beq
\psi_{\lambda P}(x_1, x_2)
\es 
\langle 0 | \psi^{+}_{1\lambda}(x_1) \, \psi^{+}_{2\lambda}(x_2) 
{1 \over P^- - H_{\lambda 0} } \,  
\, 1_{12} \, H_{\lambda I} |P\rangle \, .
\eeq
The result takes the form 
\beq
\label{wf}
\psi_{\lambda P}(x_1, x_2)
\es
\int d^3x_1' \, d^3x_2' \,
G_{\lambda P}(x_1, x_2 ; x_1',x_2') \,
\phi_{\lambda P}(x_1', x_2') \, ,
\eeq
in which the matrix element 
\beq
G_{\lambda P}(x_1, x_2 ; x_1',x_2')
\es
\langle 0 | \psi^{+}_{1\lambda}(x_1) \, \psi^{+}_{2\lambda}(x_2) 
{1 \over P^- - H_{\lambda 0} } 
\psi_{1\lambda}(x_1') \, \psi_{2\lambda}(x_2') \, |0
\rangle \, ,
\nn
\eeq
appears in the role of a two-body Green's function 
for states with LF energy $P^-$ on the LF hyperplane 
$x^+=0$. The matrix element of Eq. (\ref{vf})
plays the role of a vertex function in Eq. (\ref{wf}).

The Hamiltonian $H_{\lambda I}$ from Eq.
(\ref{HlambdaI5}), plainly modified to the case of 
three species of particles by associating position
$x_k$ with species $k$ for $k = 1, 2, 3$ and keeping 
the factor 3 in front, can be inserted into 
Eq. (\ref{vf}), which yields
\beq
\label{vf2}
\phi_{\lambda P}(x_1, x_2) \es 
3 g_\lambda \, 
\int_0^1 {dz P^+ \over 2 (2\pi)} \int {d^2 q^\perp \over (2\pi)^2 } 
\, f_\lambda \,
e^{ - i \left\{ (zP+q) x_1 + [(1-z)P-q] x_2
\right\} } \, .
\nn
\eeq
By comparison with Eq. (\ref{barf}), the following 
relation is uncovered between the non-local  
interaction Hamiltonian densities obtained in 
first-order RGPEP in previous sections and a 
2-body vertex function on the LF:
\beq
%\label{barf}
g_\lambda \bar f_\lambda(x_1, x_2, x_3)
\es
\int_0^\infty {dP^+       \over 2 (2\pi)   } 
\int          {d^2P^\perp \over   (2\pi)^2 } 
\, \phi_{\lambda P}(x_1, x_2)\, e^{iPx_3 } \, .
\eeq
Inverting the Fourier transform, one obtains
\beq
\label{phibarf}
\phi_{\lambda P}(x_1, x_2)
\es
g_\lambda \int d^3x_3 \, 
\bar f_\lambda(x_1, x_2, x_3) \, e^{-iPx_3 } \, .
\eeq
Using expressions derived for $\bar f_\lambda(x_1,x_2,x_3)$
in previous sections, one obtains in the case $m=0$ 
\beq
\label{masslessvertex}
\phi_{\lambda P}(x_1, x_2)
\es
3g_\lambda \, \left( {\lambda \over 4 \pi } \right)^2 \, P^+ \, e^{ - i PR } 
\nt
\int_0^1 dz \, z(1-z) \,
e^{-i(z-1/2) Pr - {1 \over 4} z(1-z) \, \lambda^2
r^{\perp 2} } \, ,
\eeq
and in the case $\lambda \lesssim m$ 
\beq
\label{massivevertex}
\phi_{\lambda P}(x_1, x_2)
\es
3g_\lambda \, \left( {\lambda \over 4 \pi } \right)^2 \, P^+ \, e^{ - i PR } 
\nt
C(\lambda/m) 
\quad
e^{- {\lambda^4 \over 96 m^2} \left[ \left( {Pr \over 2m} \right)^2 +  
r^{\perp 2} \right] } \, ,
\eeq
where 
\beq
C(\lambda/m) \es 
e^{- 9m^4/\lambda^4} \, 
{\lambda^4 \over 96 \, m^4} \, 
\sqrt{ \pi \over 6} \, 
 \, .
\eeq

Irrespective of the values of particle masses,
RGPEP scale $\lambda$, and the coupling constant
$g_\lambda$, the vertex function contains the
factor $P^+ \lambda^2$ which carries its
dimension. The vertex function contains as a
factor a plane-wave function of the center-of-mass
position variable $R = (x_1 + x_2)/2$. Particles 1
and 2 contribute equally to $R$ because $m_1 = m_2
= m$. The remaining factor is a function of
relative motion but depends on the total momentum
$P$. More precisely, it has a universal property
of being a function of two variables: square of
$r^\perp = x_1^\perp - x_2^\perp$ and $Pr$. Taking
into account that $r^+ = 0$, the former variable
is equal to a square of a four-vector, $-r^2$, and
the latter variable is a product of two
four-vectors. Both variables are invariant under 7
kinematical LF symmetries.

The generic structure of the vertex function
implies that the Soper variable $R_s = z x_1 +
(1-z) x_2$ \cite{Soper} does not properly separate
the center-of-mass motion from the relative motion
of constituents. Namely, the parameterization $p_1
= z P + k$ and $p_2 = (1-z) P - k$ implies $p_1
x_1 + p_2 x_2 = PR_s + kr$, where the relative
momentum $k = (1-z) p_1 - z p_2$ has $k^+ \equiv
0$. This means that the product $PR_s$ is a
mixture of $PR$ and $Pr$ that depends on $z$. $Pr$
and $r^\perp$ can be arguments of the
relative-motion vertex function, but $z$ is not
allowed to appear in the plane wave that describes
the center-of-mass motion with a definite total
momentum $P$. The product $Pr$ is a natural
variable to complement $r^\perp$ as an argument of
the vertex function even though these variables
have different dimensions (see below).

Another reason for the found structure of the
vertex function to be of interest is that there
exists an analogy between the LF wave function and
AdS/CFT descriptions of bound-state form factors,
discovered by Brodsky and de T\'eramond
\cite{BrodskyTeramond}. In that analogy, the
effective transverse distance variable $
\zeta^\perp = \sqrt{z(1-z)} \, r^\perp$ plays a
key role as an argument of a bound-state wave
function. The question concerning
$\lambda$-dependent non-local LF interaction
Hamiltonian densities is whether their RGPEP
evolution can be understood as dependence on a 5th
dimension \cite{APP5thdimension} in the context of
duality \cite{PolyakovWall, stringsQCD,
MaldacenaReport} and whether this dependence can
explain the shape of wave functions found by
Brodsky and de T\'eramond. They interpret the
analogy between AdS/CFT duality and AdS/QCD
picture of hadrons in LF formulation without any
need for considering the argument $Pr$ and scale
$\lambda$.

The issue of different dimensions of $Pr$ and
$r^2$ is resolved in the large-$\lambda$ case,
i.e., $\lambda \gg m \rightarrow 0$ and Eq.
(\ref{masslessvertex}), by multiplication of
$r^\perp$ by $\lambda$. In the small-$\lambda$
case, i.e., $\lambda \lesssim m$ and Eq.
(\ref{massivevertex}), the same issue is resolved
by dividing $Pr$ by the sum of masses of the
interacting constituents. The ratio of $\lambda$
to the masses becomes a dimensionless parameter.

It is worth noting that the relativistic Eq.
(\ref{massivevertex}) predicts in the CMS a small
difference between how $r^\perp=(r_x,r_y)$ and
$r_z$ enter the vertex function. Namely, for
$P^\perp=0$, the ratio $Pr/(2m)$ becomes $P^+
r^-/(4m)$ and $P^+$ in the CMS equals mass, say
$M$, of the state under consideration (consider
the Lee model \cite{LeeModel}), while $r^- = -2
r^3$ when $r^+=0$. So, $[Pr/(2m)]^2$ in Eq.
(\ref{massivevertex}) becomes $[Mr_z/(2m)]^2$, or
$[1-E_B/(2m)]^2 r_z^2$, where $E_B$ denotes
binding energy. This is how the LF vertex function
keeps track of the mass defect due to binding.
This result requires better understanding than the
one offered here. On the other hand, the same Eq.
(\ref{massivevertex}) can be used in any frame,
including the IMF, where one sees how the
interaction vertex or wave function get squeezed
due to motion. This is a relativistic squeezing 
in a quantum theory, not classical.

Finally, it also seems worth mentioning that the
relationship identified here between non-local LF
interaction vertices and bound-state vertex
functions may become helpful in calculating
observables such as form factors. The suggestion
is based on the fact that the function $\bar
f_\lambda (x_1, x_2, x_3)$ defined in Eq.
(\ref{barf}) gives the same interaction
Hamiltonian in Eq. (\ref{HlambdaI6}) that is also
obtained in Eq. (\ref{HlambdaIp}) using $\tilde
f_\lambda(x_1, x_2, x_3)$ defined in Eq.
(\ref{tildef}). Suppose that the old-fashioned
perturbation theory for form factors
\cite{formfactors} can be developed using
non-local interaction Hamiltonians of the type
defined in Eq. (\ref{HlambdaIp}) with non-locality
of the type defined in Eq. (\ref{tildef}). Feynman
rules with non-local vertices could then suggest
how to incorporate all diagrams that count for all
kinds of momentum transfers, not only those that
have $q^+=0$. In the case of local vertex
functions \cite{GlazekSawicki}, it is known what
to do when $q^+\neq 0$. It is less clear what to
do for bound-state vertices that involve
non-trivial vertex functions. Therefore, it seems
worth checking if the non-local interaction
Hamiltonian densities obtained in first-order
RGPEP lead to unique answers.

%%%%%%%%%%%%%%%%%%%%%%%%%%%%%%%%%%%%%%%%%%%%%%%%%%%%%%%%%%%%
%%%%%%%%%%%%%%%%%%%%%%%  CONCLUSION  %%%%%%%%%%%%%%%%%%%%%%%
%%%%%%%%%%%%%%%%%%%%%%%%%%%%%%%%%%%%%%%%%%%%%%%%%%%%%%%%%%%%

\section{ Conclusion }
\label{c}

Renormalized Hamiltonian densities on the LF
hyperplane in space-time contain interaction terms
that are non-local. The non-locality is certainly
intriguing and needs to be studied for many
reasons as a feature of basic interactions. This
article makes only a first step in this direction.
In particular, we calculate the non-locality in
first-order of a perturbative expansion in powers
of an effective coupling constant using RGPEP.
This is done for terms that originate from a
product of three fields. Such terms include
coupling of fermions to gauge bosons, coupling of
fermions to Yukawa particles, and coupling of
non-Abelian gauge bosons with themselves. But the
leading non-locality is common to all these cases
and can be calculated using scalar fields. The
result is that the non-local interaction density
has a generic form as a function of the space-time
positions of effective particles that are created
and annihilated by the interaction. This form can
be understood in terms of a vertex function for a
two-body bound state. 

The characteristic dependence of the vertex
function on $Pr$ and $r^2$, where $P$ denotes the
bound-state total momentum and $r$ denotes the
relative position of its two constituents, implies
a characteristic dependence of the non-local
Hamiltonian density on the position $x_3$ of an
annihilated (created) particle relatively to the
positions $x_1$ and $x_2$ of created (annihilated)
particles in a three-particle vertex. Namely,
$x_3$ is distributed near a straight line
connecting $x_1$ with $x_2$ on the LF. Figs. 1 to
5 illustrate the shapes of the calculated
distributions. 

All examples studied here were obtained for all
particles having the same mass. For different
masses, the results will be numerically different
and of more interest from the point of view of
application. However, there is no reason to expect
a major alteration in the method and results
beyond numerical changes. Nevertheless, such
changes will be significant in practice.

Another need for generalization concerns
interactions that originate from products of more
than three fields in one vertex. RGPEP provides a
set of general rules for how to calculate
non-local interaction densities in such cases
order-by-order in perturbation theory. It is not
excluded that the terms identified using RGPEP in
perturbation theory will lead to a selection of
dominant terms for which a non-perturbative
evolution can be derived on a computer.

A general speculation is in order concerning the
role of non-local interactions in formation of
strings of quantum gluons. Imagine that an
effective gluon splits non-locally into two along
a line as required by the LF non-locality of a
Hamiltonian interaction term. Then, each of the
two gluons interacts with neighboring gluons. It
was argued before using analogy with RGPEP results
for heavy quarkonia \cite{strings} that gluons of
small $\lambda$ may attract each other in color
singlets by potentials that resemble harmonic
oscillator. In this context, the non-local
splitting of gluons can be seen as a candidate for
dynamical generation of quantum strings in which
interactions between neighboring gluons in space
are so strong that the string energy grows only
linearly with its length, the effective gluon mass
at scale $\lambda$ providing a unit of energy per
unit of length order $1/\lambda$. 

\newpage

%%%%%%%%%%%%%%%%%%%%%%%%%%%%%%%%%%%%%%%%%%%%%%%%%%%%%%%%%%%%
%%%%%%%%%%%%%%%%%%%%%%%   APPENDIX   %%%%%%%%%%%%%%%%%%%%%%%
%%%%%%%%%%%%%%%%%%%%%%%%%%%%%%%%%%%%%%%%%%%%%%%%%%%%%%%%%%%%

\begin{appendix}

%%%%%%%%%%%%%%%%%%%%%%%%%%%%%%%%%%%%%%%%%%%%%%%%%%%%%%%%%%%%
%%%%%%%%%%%%%%%%%%%%%%%  APPENDIX A  %%%%%%%%%%%%%%%%%%%%%%%
%%%%%%%%%%%%%%%%%%%%%%%%%%%%%%%%%%%%%%%%%%%%%%%%%%%%%%%%%%%%

\section{ Connection of Eqs. (\ref{HlambdaI6}) 
and (\ref{barf}) with Eqs. (\ref{HlambdaIp}) and
(\ref{tildef})}
\label{a:connection}

Details of the connection of interest are provided
here for completeness. The connection involves
several steps. Each of them involves manipulation
of several variables. Understanding the connection
requires tracing of these steps. These steps also
exhibit an analogy between the momentum labeling
of creation and annihilation operators in RGPEP
and the parameters $\eta$ in the range between 0
and 1 that appear in old-fashioned perturbation
theory for scattering processes in the IMF
\cite{Weinberg}, or parameters $x$ in the same
range from 0 to 1 that appear in $x^+$-ordered
Feynman rules for calculating scattering
amplitudes~\cite{Yan2}. 

Creation and annihilation operators in the interaction 
Hamiltonian in Eq. (\ref{HlambdaI5}) stand in normal 
order and inserting a sign of normal ordering does not 
change anything. Two terms, both with integration over 
$P^+>0$, can be changed to one term with integration 
over $P^+$ from $-\infty$ to $+\infty$, rendering the 
result that $H_{\lambda I}$ in Eq. (\ref{HlambdaI6})
equals
\beq 
\label{aHlambdaI5} 
H_{\lambda I} 
\es 
3 g_\lambda \, 
\int          {d^3P \over (2\pi)^3 } 
\int_0^1 {dz |P^+| \over 2 (2\pi)} \int {d^2 q^\perp \over (2\pi)^2 } 
\, f_\lambda \,
\nt
\int d^3x_1 d^3x_2 d^3x_3 \, e^{ - i X } \, 
: \psi_\lambda(x_1)\psi_\lambda(x_2) \psi_\lambda(x_3) : \, . 
\eeq
One can use $\bar f_\lambda $ defined in Eq. (\ref{barf}) 
to introduce 
\beq
\hat f_\lambda \es \bar f_\lambda + \bar
f_\lambda^* \, , 
\eeq
and write
\beq 
\label{aHlambdaI6} 
H_{\lambda I} 
\es 
g_\lambda \,
\int d^3x_1 d^3x_2 d^3x_3 \, 
\hat f_\lambda(x_1, x_2, x_3) \, 
: \psi_\lambda(x_1)\psi_\lambda(x_2)
\psi_\lambda(x_3) : \, ,
\nn
\eeq
where 
\beq
\label{abarf}
\hat f_\lambda(x_1, x_2, x_3)
\es
3 
\int          {d^3P \over (2\pi)^3 } 
\int_0^1 {dz |P^+| \over 2 (2\pi)} \int {d^2 q^\perp \over (2\pi)^2 } 
\, f_\lambda \, e^{ - i X } \, , \\
\label{aX}
X \es (zP+q) x_1 + [(1-z)P-q] x_2 - P x_3 \, , \\
f_\lambda 
\es \exp{  \left\{ - \left[ {q^{\perp 2} + m^2 \over z(1-z) } - m^2 \right]^2 /\lambda^4   \right\} } \, .
\eeq
The function $\hat f_\lambda$ in Eq. (\ref{aHlambdaI6}), 
defined by Eq. (\ref{abarf}), should be the same 
as the function $\tilde f_\lambda$ in Eq. (\ref{HlambdaIp}), 
defined by Eq. (\ref{tildef}), i.e.,
\beq
\label{atildef}
\tilde f_\lambda(x_1, x_2 , x_3)
\es
\left[ \prod_{i=1}^3 
\int { d^3 p_i \over 2 (2\pi)^3} \right]  \,
2 (2\pi)^3 \, \delta^3(p_1 + p_2 + p_3) 
\nt
e^{- (\Delta{\cal M}^2/\lambda^2  )^2}  \, 
e^{ + i (p_1 x_1 + p_2 x_2 + p_3 x_3) } \, .
\eeq
In order to exhibit equivalence of $\tilde f_\lambda$ 
in Eq. (\ref{atildef}) and $\hat f_\lambda$ in Eq.
(\ref{abarf}), we change integration variables
in Eq. (\ref{atildef}) according to Eqs. (\ref{p1standard}) 
and (\ref{p2standard}), from $p_1$ and $p_2$ to 
$P_{12}$, $\zeta$ and $\kappa$,
\beq
p_1^+     \es \zeta P^+_{12} \, , 
\quad
p_1^\perp \rs \zeta P^\perp_{12} + \kappa^\perp \, , 
\quad
p_1^-     \rs { p_1^{\perp 2} + m^2 \over \zeta P^+_{12} } \, , \\
p_2^+     \es (1-\zeta) P^+_{12} \, , 
\quad
p_2^\perp \rs (1-\zeta) P^\perp_{12} - \kappa^\perp \, , 
\,
p_2^-     \rs { p_2^{\perp 2} + m^2 \over (1-\zeta)P^+_{12} } \, , 
\eeq
rename $p_3$ to $P$, and obtain
\beq
\label{atildef2}
\tilde f_\lambda(x_1, x_2 , x_3)
\es
\int { d^3 P                                \over 2 (2\pi)^3} 
\int { d^3 P_{12}                           \over 2 (2\pi)^3} 
\int_{-\infty}^{+\infty} {d\zeta |P^+_{12}| \over 2 (2\pi)  } 
\int {d^2\kappa^\perp                       \over   (2\pi)^2} 
\nt
2 (2\pi)^3 \, \delta^3(P_{12} + P) \,
e^{- (\Delta{\cal M}^2/\lambda^2  )^2}  \, e^{ - i Y }
\, , \\
Y
\es
- (\zeta P_{12}+\kappa) x_1 - [(1-\zeta) P_{12} -
\kappa] x_2 - P x_3 \, ,
\eeq
with $\Delta {\cal M}^2$ given by Eqs. (\ref{masses}) and (\ref{momenta}).
Change of the variable $\kappa^\perp$ to $-\kappa^\perp$ and 
integration over $P_{12}$ yields
\beq
\label{atildef3}
\tilde f_\lambda(x_1, x_2 , x_3)
\es
\int { d^3 P                                \over 2 (2\pi)^3} 
\int_{-\infty}^{+\infty} {d\zeta |P^+|      \over 2 (2\pi)  } 
\int {d^2\kappa^\perp                       \over   (2\pi)^2} 
\,
e^{- (\Delta{\cal M}^2/\lambda^2  )^2}  \, e^{ - i Y } , 
\nn
\\
Y
\es
(\zeta P + \kappa) x_1 + [(1-\zeta) P - \kappa] x_2 - P x_3 \, ,
\eeq
which is to be compared with Eq. (\ref{abarf})
for $\hat f_\lambda(x_1,x_2,x_3)$ using
\beq
\label{ap1}
p_1^+     \es -\zeta P^+ \, , 
\quad
p_1^\perp \rs -\zeta P^\perp - \kappa^\perp \, , 
\quad
p_1^-     \rs { p_1^{\perp 2} + m^2 \over -\zeta P^+ } \, , \\
\label{ap2}
p_2^+     \es -(1-\zeta) P^+ \, , 
\quad
p_2^\perp \rs -(1-\zeta) P^\perp + \kappa^\perp \, , 
\quad
p_2^-     \rs { p_2^{\perp 2} + m^2 \over -(1-\zeta)P^+ } \, , 
\nn \\
\label{ap3}
p_3^+     \es P^+ \, , 
\quad
p_3^\perp \rs P^\perp \, , 
\quad
p_3^-     \rs { P^{\perp 2} + m^2 \over P^+ } \, ,
\eeq
and also Eqs. (\ref{masses}) and (\ref{momenta}).

Note that Eqs. (\ref{atildef3}) and (\ref{abarf})
look similar in terms of integrations and
functions they involve. For example, both involve
the same integration over $P$ and the functions
$X$ and $Y$ coincide after replacement of $z$ and
$q^\perp$ by $\zeta$ and $\kappa^\perp$,
respectively. However, the two expressions as a
whole differ by a factor of 3 in front. The
integration over $z$ ranges from 0 to 1 while
integration over $\zeta$ ranges from $-\infty$ to
$+\infty$. This difference is analogous to the
difference between parameters $\eta$~\cite{Weinberg} 
or parameters $x$~\cite{Yan2} that range from 0 to 1 
($\eta$ and $x$ were mentioned at the beginning of 
this Appendix) and the usual momentum variables that 
range from $-\infty$ to $+\infty$. The form factor 
$f_\lambda$ appears defined in terms of particle 
momenta differently in both cases. In order to see 
how these differences conspire to produce the same 
result, we first evaluate $\Delta {\cal M}^2$ in Eq.
(\ref{atildef3}).

We divide the range of integration over $\zeta$ in
Eq. (\ref{atildef3}) into three ranges; one from 
$-\infty$ to 0, shortly called 1, one from 
0 to 1, called 2, and one from 1 to $\infty$, 
called 3. In each of these 3 regions, we change 
variables of integration from $\zeta$ and
$\kappa^\perp$ to new variables and demonstrate
that the resulting expression coincides with 1/3
of $\hat f_\lambda(x_1,x_2,x_3)$ in Eq.
(\ref{abarf}).

Eqs. (\ref{masses}) and (\ref{momenta}) imply
\beq
\Delta {\cal M}^2 \es
(p_1 + p_2 + p_3)(s_1 p_1 + s_2 p_2 + s_3 p_3) \, ,
\eeq
where $s_i = {\rm sgn}(p_i^+)$. Using Eqs.
(\ref{ap1}) to (\ref{ap3}),
\beq
s_3 \es s_P \, ,
\quad
s_2 \rs - s_{1-\zeta} s_P \, , 
\quad
s_1 \rs - s_\zeta s_P \, .
\eeq
Thus, for all particles in the interaction term 
having the same mass $m$,
\beq
\label{deltam}
\Delta {\cal M}^2 s_P 
\es
   m^2 (1 - s_\zeta - s_{1-\zeta}) 
\nn
\ps
(1 - s_\zeta)     p_1 p_3
+
(1 - s_{1-\zeta}) p_2 p_3
-
(s_\zeta + s_{1-\zeta}) p_1 p_2 
\, .
\eeq
The resulting four-vector products read
\beq
p_1 p_2 
\es
{ \kappa^{\perp 2} + m^2 \over 2 \zeta(1-\zeta)} -
m^2 \, , \\
p_1p_3
\es
- { \zeta \, m^2 \over 2} - { \kappa^{\perp 2} + m^2 \over
2 \zeta }  \, , \\
p_2p_3
\es
-{(1-\zeta) \, m^2 \over 2} - { \kappa^{\perp 2} + m^2
\over 2(1-\zeta) } \, ,
\eeq
and the result is that 
\beq
\label{atildef7}
\tilde f_\lambda(x_1, x_2 , x_3)
\es
\int { d^3 P                                \over 2 (2\pi)^3} 
\int_{-\infty}^{+\infty} {d\zeta |P^+|      \over 2 (2\pi)  } 
\int {d^2\kappa^\perp                       \over   (2\pi)^2} 
\,
e^{- (\Delta{\cal M}^2/\lambda^2  )^2}  \, e^{ - i Y } , 
\nn
\\
\Delta {\cal M}^2 
\es
\left[ m^2 - { \kappa^{\perp 2} + m^2 \over
\zeta(1-\zeta) } \right] 
{1 + |\zeta| + |1-\zeta| \over 2 s_P}  \, , \\
\label{aY7}
Y
\es
(\zeta P + \kappa) x_1 + [(1-\zeta) P - \kappa] x_2 - P x_3 \, ,
\eeq
should match the corresponding expression in
Eq. (\ref{abarf}), repeated here for the readers'
convenience,
\beq
\label{abarf7}
\hat f_\lambda(x_1, x_2, x_3)
\es
3 
\int          {d^3P \over (2\pi)^3 } 
\int_0^1 {dz |P^+| \over 2 (2\pi)} \int {d^2 q^\perp \over (2\pi)^2 } 
\, e^{- (\Delta{\cal M}^2/\lambda^2  )^2} \, e^{ -
i X } \, , \nn
\\
\Delta {\cal M}^2 
\es {q^{\perp 2} + m^2 \over z(1-z) } - m^2 \, , \\
\label{aX7}
X \es (zP+q) x_1 + [(1-z)P-q] x_2 - P x_3 \, .
\eeq
The 0 to 1 part of integration over $\zeta$ 
in Eq. (\ref{atildef7}) matches exactly 1/3 
of Eq. (\ref{abarf7}). The question is how 
to see the matching of the remaining 2/3
with parts of integration over $\zeta$ from
$-\infty$ to 0 and from 1 to $+\infty$.

There are two main regions of integration
variables, $P^+ > 0$ and $P^+ < 0$. Focus 
on the region $P^+ > 0$. Split integration 
over $\zeta$ into three ranges; range 1 
from $-\infty$ to 0, range 2 from 0 to 1, 
and range 3 from 1 to $\infty$. In the region 
2, results match. Consider region 1. For 
$\zeta <0$, one has $p_1^+ > 0$, which 
means the particle 1 is annihilated, not 
created, particle 2 is created, and particle 
3 is always annihilated for $P^+>0$.

So, for $\zeta < 0$, change variables
treating $p_2$ as a total momentum (created, 
with a negative $+$ component) composed of
$p_1$ and $p_3$ (both annihilated, with positive 
$+$ components). This means 
\beq
p_1 \es - z p_2 - q \, , 
\quad 
p_3 \rs - (1-z) p_2 + q \, ,
\eeq
and
\beq
\label{ap1negative}
p_1^+     \es -\zeta P^+ \rs - z p_2^+ \, , 
\quad
p_1^\perp \rs -\zeta P^\perp - \kappa^\perp 
          \rs - z p_2^\perp - q^\perp \, , \\
\label{ap2negative}
p_2^+     \es -(1-\zeta) P^+ \rightarrow - P^+ \, , 
\quad
p_2^\perp \rs -(1-\zeta) P^\perp + \kappa^\perp \rightarrow - P^\perp \, , 
\nn \\
\label{ap3negative}
p_3^+     \es P^+ \rs - (1-z) p_2^+ \, , 
\quad
p_3^\perp \rs P^\perp \rs - (1-z)p_2^\perp + q^\perp \, . 
\eeq
The required change of variables is
\beq
\zeta \es { - z \over 1-z} \, , 
\quad 
\kappa^\perp \rs { q^\perp \over 1-z } \, , \\
d\zeta \es { - dz \over (1-z)^2 } \, , 
\quad 
d^2\kappa^\perp \rs { d^2 q^\perp \over (1-z)^2 }
\rs (1-\zeta)^2 d^2 q^\perp \, .
\eeq
With this change of variables ($s_P =1$ here but it is kept as $s_P$),
\beq
\Delta {\cal M}^2 
\es
\left[ m^2 - { \kappa^{\perp 2} + m^2 \over
\zeta(1-\zeta) } \right] 
{1 + |\zeta| + |1-\zeta| \over 2 s_P} \\
\es
\left[ { q^{\perp 2} + m^2 \over
z(1-z) } -  m^2 \right] 
{1 \over s_P} \, ,
\eeq
which is the expected function of $z$ and $q^\perp$.
So, after the change of variables, the region 1 
with $P^+>0$ contributes 
\beq
\tilde f_1(x_1, x_2 , x_3)
\es
\int { d^3 P \, \theta(P^+)         \over 2 (2\pi)^3} 
\int_{-\infty}^0 {d\zeta |P^+|      \over 2 (2\pi)  } 
\int {d^2\kappa^\perp               \over   (2\pi)^2} 
\,
e^{- (\Delta{\cal M}^2/\lambda^2  )^2}  \, e^{ - i Y } \, , \nn 
\\
\es
\int     { d^3 P \, \theta(P^+) \over 2 (2\pi)^3} 
\int_0^1 { dz |P^+|             \over 2 (2\pi)  } 
\int     { d^2q^\perp           \over   (2\pi)^2} 
\,
{e^{- (\Delta{\cal M}^2/\lambda^2  )^2} \over (1-z)^4} \, e^{ - i Y } , 
\nn
\\
Y
\es
{ - z P + q \over 1- z } \, x_1 
+ 
{ P - q \over 1 - z} \, x_2 - P x_3 \, .
\eeq
Change notation $p_2 \rightarrow -
\tilde P$, 
\beq
\label{tildeap2negative}
p_2^+     \es -(1-\zeta) P^+ \rs - \tilde P^+ \, , 
\quad
p_2^\perp \rs -(1-\zeta) P^\perp + \kappa^\perp
\rs - \tilde P^\perp \, ,
\nn
\eeq
keeping $s_P = s_{\tilde P}$, so that
\beq
P \es (1-z) \tilde P + q \, .
\eeq
Then,
\beq
\label{atildef10}
\tilde f_1(x_1, x_2 , x_3)
\es
\int     { d^3 \tilde P \,(1-z)^3 \theta(\tilde P^+) \over 2 (2\pi)^3} 
\int_0^1 { dz |\tilde P^+|(1-z)             \over 2 (2\pi)  } 
\nt
\int     { d^2q^\perp           \over   (2\pi)^2} 
\,
{e^{- (\Delta{\cal M}^2/\lambda^2  )^2} \over
(1-z)^4} \, e^{ - i Y } \, ,
\nn
\eeq
where
\beq
Y
\es
( - z \tilde P + q) \, x_1 
+ 
\tilde P \, x_2 
- 
[(1-z) \tilde P + q] x_3 \, .
\eeq
Changing variable $\tilde P$ to $-P$, 
\beq
\label{atildef12}
\tilde f_1(x_1, x_2 , x_3)
\es
\int     { d^3 P \, \theta(-P^+) \over 2 (2\pi)^3} 
\int_0^1 { dz |P^+|              \over 2 (2\pi)  } 
\int     { d^2q^\perp            \over   (2\pi)^2} 
\,
e^{- (\Delta{\cal M}^2/\lambda^2  )^2} \, e^{ - i Y } \, ,
\nn
\\
\Delta {\cal M}^2 
\es
\left[ { q^{\perp 2} + m^2 \over
z(1-z) } -  m^2 \right] 
{- 1 \over s_P} \, , \\
Y
\es
   (z P + q) \, x_1 
+ [(1-z) P - q] x_3 
- P \, x_2 \, .
\eeq
This result should be compared with 1/3 of 
$\hat f_\lambda(x_1, x_2, x_3)$ given in Eq.
(\ref{abarf7}).

This comparison shows that initial integration
range over $P^+ > 0$ and thus particles 1 and 2
created and particle 3 annihilated in $\tilde f$, 
corresponds to integration over $P^+ < 0$ and 
thus particles 1 and 3 annihilated and particle 2 
created in $\hat f$. When both signs of $P^+$ are 
included in the integration, the result must be 
\beq
\label{range1}
\tilde f_1(x_1, x_2 , x_3)
\es {1\over 3} \hat f(x_1, x_3 , x_2) \, .
\eeq
The next observation is that the integration over 
$x_2$ and $x_3$ in the Hamiltonian includes only 
the symmetric part of the functions $f(x_1, x_2,
x_3)$ in Eqs. (\ref{HlambdaIp}) and (\ref{aHlambdaI6}).
Therefore, Eq. (\ref{range1}) completes the 
explanation of how the integration over range 1 
produces 1/3 of the interaction Hamiltonian.

Since the integration in range 3 can be
transformed in the same way, the only difference
being that for $\zeta > 1$ the particle 1 and
particle 2 are changed in their roles with respect
to particle 3, it follows that the integration
over $\zeta$ in range 3 produces the remaining 1/3
of the Hamiltonian. Hence, the connection of Eqs.
(\ref{HlambdaI6}) and (\ref{barf}) with Eqs.
(\ref{HlambdaIp}) and (\ref{tildef}) is
established, and both ways of writing the
non-local interaction Hamiltonian that results
from first-order RGPEP, are equivalent.

%%%%%%%%%%%%%%%%%%%%%%%%%%%%%%%%%%%%%%%%%%%%%%%%%%%%%%%%%%%%
%%%%%%%%%%%%%%%%%%%%%%%  APPENDIX B  %%%%%%%%%%%%%%%%%%%%%%%
%%%%%%%%%%%%%%%%%%%%%%%%%%%%%%%%%%%%%%%%%%%%%%%%%%%%%%%%%%%%

\section{ Integrals involved in non-locality for $m=0$ }
\label{a:m=0}

In terms of variables
\beq
\label{R}
R \es (x_1 + x_2)/2 \, , \\
\label{r}
r \es  x_1 - x_2 \rs x/\lambda \, , 
\eeq
which imply 
\beq
\label{x1}
x_1 \es R + r/2 \, , \\
\label{x2}
x_2 \es R - r/2 \, ,
\eeq
Eq. (\ref{fm0}) reads
\beq
\bar f_\lambda(x_1, x_2 , x_3)
\es
3 \int { d^3 P \, \theta(P^+) \over 2 (2\pi)^3} 
\int_0^1 {dz \, P^+         \over 2 (2\pi)  } 
\int {d^2q^\perp            \over   (2\pi)^2} 
\,
e^{- {q^{\perp 2} \over \lambda^2 z (1-z)} } 
\nt
e^{-iP[R + (z-1/2) r - x_3] + i q^\perp r^\perp } \, .
\eeq
Integration over $q^\perp$ renders
\beq
\int {d^2q^\perp            \over   (2\pi)^2} 
\,
e^{- {q^{\perp 2} \over \lambda^2 z (1-z)} } \,
e^{i q^\perp r^\perp } 
\es
{\lambda^2 z(1-z)\pi \over   (2\pi)^2} 
\,
e^{- ( r^\perp \lambda \sqrt{z(1-z)} / 2 )^2 } 
\, ,
\eeq
and subsequent integration over $P^\perp$ gives
\beq
\bar f_\lambda(x_1, x_2 , x_3)
\es
3
\int_0^\infty { dP^+ P^+ \over 4 \pi} 
\int_0^1 {dz \over 4\pi } \,
{ \lambda^2 z(1-z) \over 4 \pi } 
\nt 
\delta^2\left[ R^\perp + (z-1/2) r^\perp - x_3^\perp \right]
\nt
e^{-iP^+ [R^- + (z-1/2) r^- - x_3^-]/2 - z(1-z) \lambda^2 r^{\perp 2}/4} \, .
\eeq
Integration over $P^+$ requires regularization.
By inserting $e^{-\epsilon P^+/2}$ under the integral
and assuming that $\epsilon \rightarrow 0_+$, one 
obtains 
\beq
\label{a:fm0}
\bar f_\lambda(x_1, x_2 , x_3)
\es
{ - 3 \lambda^2 \over 16\pi^3 } 
\int_0^1 
{dz \, z(1-z) 
\over 
\left[ zx_1^- + (1-z)x_2^- - x_3^-  - i\epsilon \right]^2 } 
\\
\ts
\delta^2\left[ z x_1^\perp + (1-z) x_2^\perp - x_3^\perp \right]
\,
e^{- z(1-z) \lambda^2 (x_1^\perp - x_2^\perp)^2/4}
\, . \nonumber 
\eeq
Hence, the function $g(\rho,\tau)$ defined in Eq. (\ref{grt}) 
is given by Eq. (\ref{explicitgrt}).

%%%%%%%%%%%%%%%%%%%%%%%%%%%%%%%%%%%%%%%%%%%%%%%%%%%%%%%%%%%%
%%%%%%%%%%%%%%%%%%%%%%%  APPENDIX C  %%%%%%%%%%%%%%%%%%%%%%%
%%%%%%%%%%%%%%%%%%%%%%%%%%%%%%%%%%%%%%%%%%%%%%%%%%%%%%%%%%%%

\section{ Integrals involved in non-locality 
          for $\lambda \lesssim m$ }
\label{a:lambda=m}

Introducing a dimensionless three-vectors 
$\vec \rho = \lambda \, \vec r$, where 
$\vec r$ is defined in Eq. (\ref{threevectorr}), 
and $\vec p = \vec k /\lambda$, where $\vec k$ 
is defined in Eqs. (\ref{kp}) and (\ref{kz}), one 
obtains the function $\bar f$ defined in Eq. 
(\ref{fr}) in the form
\beq
\bar f
\es
3 \, e^{- 9 m^4/\lambda^4 } \,
\int {d^3P P^+ \theta(P^+) \over 2 (2\pi)^3} \, 
e^{ - i P(R-x_3) } 
\nt
\int { \lambda^3 \, d^3p 
\over 4m (2\pi)^3 }
\,
e^{ - {24m^2 \over \lambda^2} \, \vec p\,^2 } 
\,\,
e^{ i \vec p \, \vec \rho } \, ,
\eeq
Using $ \beta^2 = 24 m^2 /\lambda^2$, one can 
perform integration over $\vec p$, obtaining
\beq
\bar f
\es
3\,
e^{- 9m^4/\lambda^4} \,
\int {d^3P P^+ \theta(P^+) \over 2 (2\pi)^3} \, 
e^{ - i P(R-x_3) } 
\nt
{\lambda^3 \over \beta^3}
{ \sqrt{\pi} ^3 
\over 4m (2\pi)^3 }
\,
e^{- \vec \rho\,^2 /(2\beta)^2 } \, .
\eeq
To integrate over $P^\perp$, $\bar f$ can 
be written in the form
\beq
\bar f
\es
3 \,e^{ - 9 m^4 / \lambda^4 } \,
{\lambda^3 \over \beta^3}
{ \sqrt{\pi} ^3 
\over 4m (2\pi)^3 }
\,
e^{- \left( {\rho^\perp \over 2\beta} \right)^2 } \, 
\int_0^\infty {dP^+ \, P^+ 
e^{ - i P^+(R^- - x^-_3)/2 } 
\over 4\pi} 
\nt
\int {d^2P^\perp \over (2\pi)^2} \,
e^{i P^\perp (R^\perp - x_3^\perp) } 
e^{- \left( {P^+ \lambda r^-/2 - P^\perp
\rho^\perp \over 4m\beta} \right)^2 } \, .
\eeq
Then, $P^\perp$ can be written in terms of 
two mutually orthogonal transverse vectors, 
$e_\rho = \rho^\perp/|\rho^\perp|$ and 
$\iota^\perp$, as
\beq
P^\perp \es (  p\, e_\rho^\perp + q \, \iota^\perp ) \lambda \, .
\eeq
The auxiliary parameter $p$ has nothing 
to do with $\vec p = \vec k/\lambda $ introduced 
earlier. The integral over $P^\perp$ becomes
\beq
& & 
\int {d^2P^\perp \over (2\pi)^2} \,
e^{i P^\perp (R^\perp - x_3^\perp) } 
e^{- \left( {P^+ \lambda r^-/2 - P^\perp \rho^\perp \over 4m\beta} \right)^2 } 
\\
\es
\lambda^2 \delta\left[ \lambda \iota^\perp ( R - x_3 )^\perp \right]
\,
\int {dp \over 2\pi } \,
e^{i p \, e_\rho^\perp (R - x_3)^\perp \lambda }
e^{- \left( {P^+ \lambda r^-/2 - p \lambda |\rho^\perp| \over 4m\beta}
\right)^2 } \, .
\eeq
The remaining integral over $p$, using $\chi =
e_\rho^\perp(R^\perp - x_3^\perp)\lambda$, gives
\beq
&& 
\int {dp \over 2\pi } \,
e^{i p \, e_r^\perp (R - x_3)^\perp \lambda }
e^{- \left( {P^+ \lambda r^-/2 - p \lambda |\rho^\perp| \over 4m\beta}
\right)^2 }  \\
\es
{4m\beta \over \lambda |\rho^\perp| } \,
e^{i {P^+r^-\chi \over 2 |\rho^\perp| }  } \,
e^{- \left( {2m\beta \chi \over \lambda |\rho^\perp| } \right)^2  }
{ \sqrt{\pi} \over 2\pi } \, .
\eeq
The entire integral is then
\beq
\bar f
\es
3 \,e^{ - 9 m^4 / \lambda^4 } \,
{\lambda^4 \over (4 \pi \beta )^2 |\rho^\perp| }
\,
e^{- \left( { |\rho^\perp| \over 2\beta} \right)^2 
   - \left( {m \chi \over \lambda } \right)^2 \, 
            \left( {2 \beta \over |\rho^\perp| } \right)^2  }
\,
\delta\left[ \lambda \iota^\perp ( R - x_3 )^\perp \right]
\nt
\int_0^\infty {dP^+ \, P^+ 
e^{ - i P^+(R^- - x^-_3)/2 } 
\over 4\pi} 
\,
e^{i {P^+r^-\chi \over 2 |\rho^\perp| }  } 
\, .
\eeq
The remaining integral over $P^+$, after the 
same regularization by factor $e^{-\epsilon P^+/2}$
that was introduced in Appendix \ref{a:m=0},
produces
\beq
\bar f
\es
3 \,
{e^{ - 9 m^4 / \lambda^4 } \, \lambda^4 
    \over 16 \pi^3 \beta^2 |\rho^\perp| }
\,
e^{- \left( { |\rho^\perp| \over 2\beta} \right)^2 
   - \left( {m \chi \over \lambda } \right)^2 \, 
            \left( {2 \beta \over |\rho^\perp| } \right)^2  }
\,
\delta\left[ \lambda \iota^\perp ( R - x_3 )^\perp
\right] 
\nt
{ -1 \over \left[ 
R^- - x^-_3 - { r^\perp (R^\perp -
x_3^\perp) \over r^{\perp 2} } \, r^-  - i \epsilon \right]^2 }
\, ,
\eeq
Using dimensionless variables $\rho = \lambda \, (x_1 - x_2)$
and $\tau = \lambda \, (x_3 - x_2)$, one can write
\beq
\chi \es 
\rho^\perp (\rho^\perp/2 -
\tau^\perp ) /|\rho^\perp| \, .
\eeq
Since $\tau^\perp$ must lie along $\rho^\perp$, so 
that $\tau^\perp = t \rho^\perp$, one gets 
\beq
\chi \es 
-(t -1/2) \, |\rho^\perp| \, . 
\eeq
Thus, the approximate result for $\bar f_\lambda(x_1, x_2, x_3) 
= \lambda^6 \, g(\rho,\tau)$, where
$\tau^\perp = t \rho^\perp + s n^\perp$, is
\beq
g(\rho,\tau)
\es
\delta (s)
\,
{3 \over 16 \pi^3} \, 
{ - 1/4 \over 
\left( \tau^- - t \rho^-  + i \epsilon \right)^2 }
\nt
{ e^{- { \rho^{\perp 2} \over 4\beta^2}  
   - { 4 \beta^2 m^2 \over \lambda^2} \, 
            \left( t - 1/2 \right)^2  }
\over |\rho^\perp| }
\,\,
{4\, e^{ - 9 m^4 / \lambda^4 } \, 
    \over \beta^2  }
\, .
\eeq
Writing 
\beq
g(\rho,\tau)
\es
\delta (s)
\,
{3 \over 16 \pi^3} 
\, 
h_\rho(t, \tau^-) \, ,
\eeq
one arrives at 
\beq
\label{ahm}
h_\rho(t, \tau^-)
\es
{ - 1/4 \over 
\left( \tau^- - t \rho^-  + i \epsilon \right)^2 }
\,
{ e^{- { \rho^{\perp 2} \over 4\beta^2}  
   - { 4 \beta^2 m^2 \over \lambda^2} \, 
            \left( t - 1/2 \right)^2  }
\over |\rho^\perp| }
\,\,
{4\, e^{ - 9 m^4 / \lambda^4 } \, 
    \over \beta^2  }
\, , \nn
\eeq
which results in Eq. (\ref{hm}).

\end{appendix}

%%%%%%%%%%%%%%%%%%%%%%%%%%%%%%%%%%%%%%%%%%%%%%%%%%%%%%%%%%%%
%%%%%%%%%%%%%%%%%%%%%%%  REFERENCES  %%%%%%%%%%%%%%%%%%%%%%%
%%%%%%%%%%%%%%%%%%%%%%%%%%%%%%%%%%%%%%%%%%%%%%%%%%%%%%%%%%%%

\end{document}